\documentclass[a4paper,11pt]{article}
\pdfoutput=1
\usepackage{jinstpub}
\usepackage{url}
\usepackage[labelformat=simple]{subcaption}

\usepackage{siunitx}
\DeclareSIUnit\sq{\ensuremath\Box}

\title{Studies on Electrical Properties of Resistive Plate Chamber (RPC) }

\author[a, b]{Subhendu Das,}
\author[a, b, 1, 2]{Jaydeep Datta,\note{Corresponding author} \note{Currently affiliated to Université Libre de Bruxelles, Av. Franklin Roosevelt 50, 1050 Bruxelles, Belgium }}
\author[a, b]{Nayana Majumdar,}
\author[a, b]{and Supratik Mukhopadhyay}

\affiliation[a]{Saha Institute of Nuclear Physics, AF Block, Sector 1, Bidhannagar, Kolkata 700064, India}
\affiliation[b]{Homi Bhabha National Institute, Training School Complex, Anushaktinagar, Mumbai 400094, India}

\emailAdd{jaydeep.datta@gmail.com}

\abstract{A numerical model based on Finite Element Method (FEM) has been developed to simulate the electrical properties of RPC utilizing the law of current conservation.
It has been used to systematically investigate the effect of the electrical as well as the geometrical parameters of the device components on the potential distribution and field configuration which govern the RPC performance.
The numerical model has been validated by comparing the dark current and electric field with those produced by another mathematical model based on surface currents.
The efficacy of the present model has been demonstrated by comparing its results to the experimental measurements performed with a glass RPC.
The measured and estimated values of these observables have been found to be in good agreement.}

\arxivnumber{2204.12986} 
\begin{document}
\maketitle
\flushbottom
\section{\label{intro} Introduction}
\label{sec:1}
\begin{figure}[h!]
    \centering
   	\includegraphics[width=0.7\textwidth]{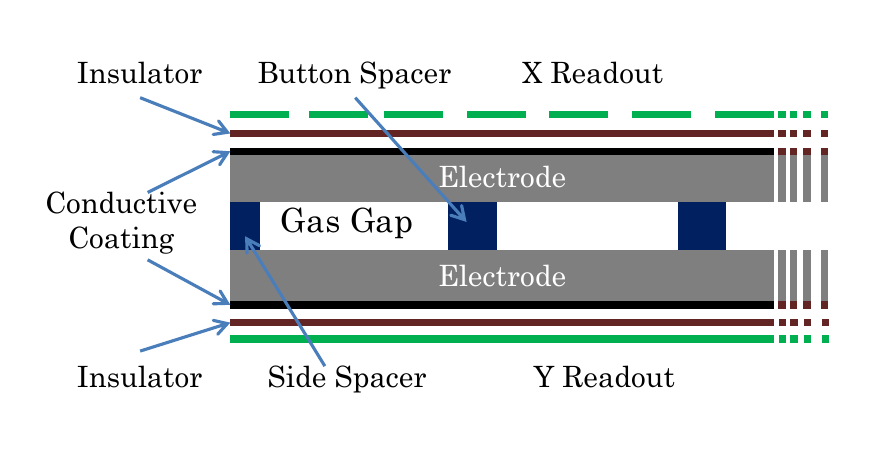}
   	\caption{Schematic RPC geometry}
    \label{fig:1}
\end{figure}

The Resistive Plate Chamber (RPC) is a gaseous particle detector introduced in the 1980s \cite{Santonico1981} as a simplified manifestation of spark chamber with localized discharge \cite{Pestov1978}.
It utilized the advantages of operation in atmospheric pressure and use of a material made of paper and phenolic resin, commonly known as Bakelite, in construction of the electrodes.
The RPC is made up of two plates of high bulk resistivity (typically glass or Bakelite), placed parallel to each other, as shown in figure \ref{fig:1}.
Depending upon the experimental requirement, they can be of $\sim 100 cm^2$ to few $m^2$ and thickness of few mm.
(The broken lines and sections on the right indicate that the RPC can be extended in that direction.)
The gap between the electrodes, usually a few mm, is filled with a gas mixture, which is confined using side spacers all around.
To maintain uniform parallel plate configuration, several small button shaped spacers are placed periodically inside the gas gap. 
Their number depends upon the dimension of the electrodes.
The spacer components (side and button) are usually made up of material (typically polycarbonate) with relative permittivity and resistivity higher than that of the electrodes.
The electrodes are coated with a conductive layer with suitably lower resistivity for the distribution of high voltage over the entire
surface.
It enables them to propagate the high voltage uniformly and at the same time become transparent to the induction of current signal on the external readout planes. 
The readout planes, made of copper strips or pads placed on firm insulating materials, are placed on either sides of the RPC and aligned in perpendicular direction to each other.
The application of the polarizing potential to the resistive electrodes via the conductive layer allows operating both the readout planes at ground potential, as they are insulated from the resistive electrodes by using a non-conducting film in between.

When a charged particle passes through the RPC, it ionizes the gas molecules.
The primary electrons can grow into an avalanche following cascaded secondary ionization processes that take place subsequently under the action of high electric field in the gas gap.
In case of very high applied field, the avalanche can become large enough to have a field with magnitude equivalent to the applied one, which leads to streamer formation \cite{deng}.
Depending upon the experimental requirements, RPC can be operated in both avalanche and streamer mode \cite{abbrescia}.
In both cases, the induced current due to the movement of the charges remains confined in a limited area (few cm$^2$) due to the high resistivity of the electrodes.
As a result, the electric field in that region drops, while the sensitivity in the rest of the chamber volume remains unaffected.
This turns out certainly useful in handling high particle flux.
The collisions between the electrons and the neutral gas molecules excite the gas molecules.
The photons emitted during the de-excitation of the electrons can propagate in the gas volume and cause secondary avalanches at other locations.
A quenching component is used in the gas mixture to hinder this process through absorption of these photons.
The propagation of the charged particles induces a current signal via capacitive coupling to the readout planes, which are placed outside the resistive electrodes.
The readout planes are insulated from the electrodes using a non-conducting layer (typically mylar).
This arrangement makes them fully independent of the active detection medium and provides a leverage for tuning the position resolution by changing the readout granularity.
Alongside, the RPC can offer a fine time resolution as an advantage arising out of the uniform field produced by its parallel plate configuration.
Moreover, its simple geometry allows easy production of large area detectors using inexpensive materials.
All these features finally lead to a cost-effective production of large scale, robust detectors with excellent time and position resolution.
As a result, the RPC has become a coveted choice for tracking and triggering purposes in various fundamental physics experiments \cite{Bacci:2000sj, Dusini:2003bu, ALICE:1999aa, ATLAS:1997ad, CMS:1997iti, LHCb:2001ab} as well as other societal fields, such as medical, portal and industrial imaging \cite{Baesso:2014xna, carloganu:in2p3-00642747, Basnet:2020edt}.

At lower voltages than that required for avalanche to set in, in the gas medium, the RPC acts as an electrical system with various resistive and capacitive components.
This gives rise to dark currents flowing through different components of the detector.
It can be noted that in the RPC, like any gaseous ionization detector, the electric field configuration plays a major role in determining the detector response by governing the multiplication and propagation of the primary electrons, which eventually lead to the formation of the signal.
The knowledge of the potential distribution and the dark current flowing through the device thus turns out to be important because these are the two key factors to determine the field configuration.
The device geometry as well as the electrical properties of the components play crucial roles in this connection, which has motivated many experimental studies to characterize the electrode materials \cite{Kaur:2014rfa, Naimuddin2014, Raveendra2016}.
On the other hand, in the present era of ever-increasing luminosity at colliders and as a consequence increased event rate, the requirement of higher rate capability has inspired to explore unconventional materials as electrodes \cite{CHAKRABORTY2019424, abbrescia}, which have bulk resistivity of orders of magnitude smaller than that of the typical choices of Bakelite and glass.
Therefore, a detailed investigation of the electrical characteristics of RPC is necessary to optimize and recommend its design and choice of materials for a specific application.

In this work, a 3D numerical model has been developed for simulating the electrical characteristics of RPC.
It utilizes the law of current conservation to calculate the dark current flowing through the device as a result of the application of high voltage.
The numerical model has been built using the commercially available Finite Element Method (FEM) solver, COMSOL Multiphysics \cite{comsol}. 
The following section \ref{nummod} describes the proposed numerical model.
The performance of the model has been studied by comparing its results with that obtained by using another numerical model, presented in \cite{Ammosov1997}.
Detailed information about the simulation conditions used in \cite{Ammosov1997} motivated its use as a benchmark.
In \cite{Ammosov1997}, the electric field and dark current of an RPC have been calculated for different conductive coating configurations using the Surface Charge Method (SCM).
It describes the electric field as superposition of fields due to surface charges and considers the effect of surface currents in the calculation of the electric field using the law of full current conservation.
It offers us a handle to validate some parameters, namely, the electric field in the gas gap, current in the spacer materials etc., which are difficult to measure in an experimental environment.
Following its comparison with the SCM model, the present numerical model has been used as a numerical tool to optimize the configuration and components of the RPC to be designed for a muon scattering tomography setup, which has been planned for material discrimination and will be further upgraded for inspection of civil structures \cite{Tripathy2020, Tripathy2021}.
In this context, a systematic study has been carried out using this model to evaluate the potential distribution, dark current and the field distribution of the RPC for different electrical and design parameters in section \ref{eval}.
To establish the findings, all these studies have been compared with experimental measurements performed on a glass RPC.
Although the results have been calculated for a fixed dimension of RPC, the conclusions drawn from them depend upon the intrinsic properties of the RPC geometry and materials, which are independent of the RPC dimensions and hence can be applied more broadly.
The time constant for potential build-up has also been studied in section \ref{eval} as a related electrical characteristic, which can be found relevant for experiments with high particle rate.
In the context of this numerical study, it should be mentioned that the values of the electrical properties of all the device components are not constant in reality, but rather dependent upon material grade and environment.
So are the physical parameters, which may have some deviation from the actual value.
In order to address this practical issue, the electrical as well as the physical parameters have been varied over a range to take the possible options into account and investigate their effect on the field distribution.
This will help to make decisions while building RPCs with material properties varying over a large range.
Finally, it is worth mentioning that the model can be used for producing a design and material guide for the implementation of RPCs for any application.
The conclusion based on the present work has been discussed in the last section \ref{con}.

\section{\label{nummod} Numerical Model}

A brief discussion on the 3D numerical model, developed in this work for studying the electrical characteristics of the RPC, has been furnished below.
In subsection \ref{math} the proposed numerical model has been described.
It has been followed by the subsection \ref{dark_curr} describing a study on the dark current flowing through the RPC, carried out using the present model.
The following subsection, \ref{comp} is dedicated to the comparison between the results from the proposed model and the model described in \cite{Ammosov1997}.
The comparison between these models has been concluded, describing the advantage and disadvantages of these models in section \ref{procon}.

\subsection{\label{math} Mathematical Approach}

When a high potential difference is applied across the electrodes, current starts to flow through the highly resistive electrodes and the spacers.
This is named as the dark current.
It basically governs the potential distribution on the electrodes, which in turn determines the electric field distribution in the gas gap.
In the present numerical model, the \textit{Electric Currents} module of COMSOL Multiphysics has been used.
It implements the continuity equation as shown in equation \ref{eq:1}.
Here, $\Vec{J}$ and $Q_{j,v}$ denote respectively, the total current density and the volumetric source of current.
	\begin{eqnarray}
	\Vec{\nabla} \cdot \Vec{J} &=& Q_{j,v} \label{eq:1}\\
	\Vec{J} &= &\sigma \Vec{E} + \Vec{J_e} \label{eq:2}\\
	\Vec{E} &= & - \Vec{\nabla} V \label{eq:3}
	\end{eqnarray}
As expressed in equation \ref{eq:2}, the total current density $\Vec{J}$ is considered as the sum of current due to electrical conductivity (expressed by $\sigma$), caused by the application of electric field $\Vec{E} $, and external current density $\Vec{J_e}$, arising out of the growth and propagation of the primary charges, caused by the passage of a charged particle. 
The electric field $\Vec{E}$ is obtained from the applied potential V following the equation \ref{eq:3}.
Following the scenario when there is no charged particle incident on the device, $\Vec{J_e}$ has been made zero, while the total current throughout the RPC volume  has been assumed to be conserved.
The volume and surface resistivity and relative permittivity of the component materials have been provided as inputs to the numerical simulation.
The infinite bulk resistivity of the gas in absence of any ionization process has been considered by using a very high value of about 10$^{18}$ \SI{}{\ohm cm}.
Its relative permittivity~has been set to 1.
In the present numerical work, the entire RPC model has been kept inside a volume of air.
As the relative tolerance level of the software solver has been set to 0.01 \cite{comsol}, all the results are accurate up to the second decimal place.

\subsection{\label{dark_curr}Flow of dark current}
\begin{figure}[h!]
	\begin{center}
		\subfloat[Dark current in the button spacers.\label{fig:button_dark_curr}]{%
			\includegraphics[width=0.48\textwidth]{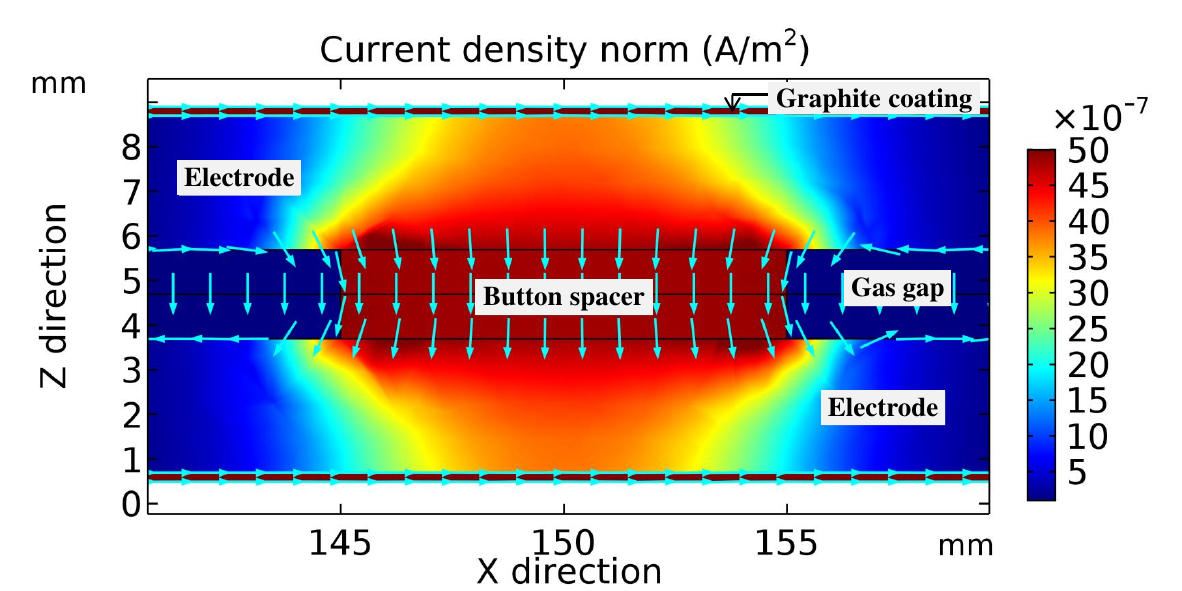}}
		\hspace{2mm}
		\subfloat[Dark current in the side spacer.\label{fig:spacer_dark_curr}]{%
			\includegraphics[width=0.48\textwidth]{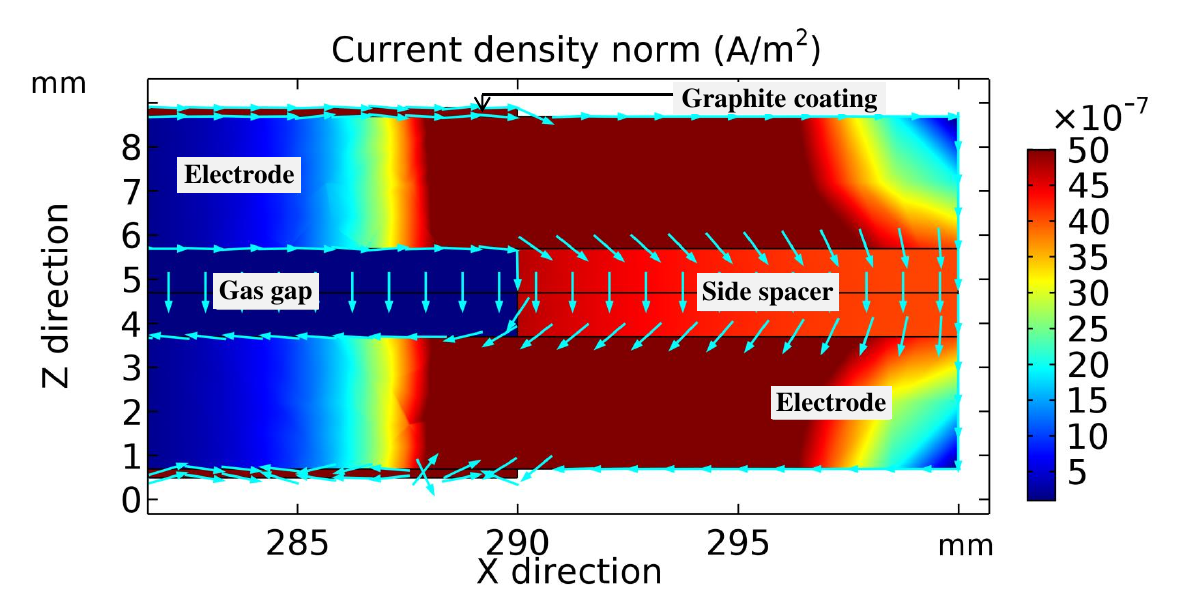}}
		\end{center}
	\caption{\label{fig:dark_curr} Dark current flowing through different parts of RPC}
\end{figure}

The simulated dark current flowing through the RPC from the numerical model has been shown in figure \ref{fig:dark_curr}.
Though the numerical model is a 3D one, the currents flowing in XZ cross-section of the RPC has been shown for clarity.
The surface current flowing in the 0.1 mm thick conductive coating is shown by the arrow heads in both the figures.
The arrow heads depict the direction of the current density, and the magnitude of the same is indicated by the color.
As the gas gap has higher volume resistivity than the spacers, most of the volume current flows through the button and side spacers as shown in the figures \ref{fig:button_dark_curr} and \ref{fig:spacer_dark_curr} respectively.
Following the continuity of volume current, a large amount flows through the electrode in contact with the spacers, which results in a drop in the applied voltage in the respective locations as discussed in section \ref{efield}.
Negligible volume current is observed in the rest of the electrode, making it transparent to the applied voltage.
To balance the high voltage applied across the gas gap, a small volume current flows in the gas gap as shown in the figures.

\subsection{\label{comp} Comparison with SCM Model}

To compare with the results produced by the SCM model \cite{Ammosov1997}, the same Bakelite RPC, used in \cite{Ammosov1997}, has been modeled in the present numerical work.
The specifications of the RPC components have been tabulated in table \ref{table1}.
Here, \textit{t, a, w, d} represent the geometrical parameters, such as thickness, area, width, diameter, while $\rho, R_s, \kappa$ denote the volume resistivity, surface resistivity and relative permittivity of the device components respectively.
Following the RPC model described in \cite{Ammosov1997}, the distance between two button spacers is 10 cm and the gap between the button and the side spacers at the edge is kept at 7.85 cm.
The conductive coat (graphite), applied all over the resistive electrode (Bakelite), has been assumed as an equipotential surface following the SCM model.
\begin{table}[h!]
\centering
\caption{\label{table1} Specifications of RPC components used in SCM model \cite{Ammosov1997}}
\begin{tabular}{ | c | c | c | c | c |  }
\hline
Components & Dimension (mm) &  $\rho$ (\SI{}{\ohm cm}) &  $R_s$ (\SI{}{\ohm \per \sq}) & $\kappa$ \\
\hline
Gas gap & $ t = 2 $ & $ 10^{18} $ & - & 1  \\
\hline
Electrode (Bakelite) & $ t = 1.6, a = 500 \times 500 $ & $ 10^{12} $ & - & 4 \\
\hline
Side spacer (PVC) & $ t = 2, w = 10 $ & $5 \times 10^{15}$ & - & 3 \\
\hline
Button spacer (PVC) & $ t =2, d =13 $ & $5 \times 10^{15}$ & - & 3 \\
\hline
Insulation film & $ t = 0.6 $ & $5 \times 10^{15}$ & - & 3 \\
\hline
Graphite coating & $ a = 500 \times 500 $ & - & $ 10^{5}$ & -\\
\hline
\end{tabular}
\end{table}
The amount of dark current flowing through different sections of the RPC obtained in this work has been compared to that produced by the SCM model.
The comparison has been presented in table \ref{table2}.
\begin{table}[h!]
\centering
\caption{\label{table2} Comparison of dark current obtained from the SCM model \cite{Ammosov1997} and this work}
\begin{tabular}{ | c | c | c | c |  }
\hline
Section of RPC & Current (nA) in \cite{Ammosov1997} & Current (nA) in this work \\
\hline
Gas gap           &   0.09    &   0.09   \\
\hline
Button spacer  &   0.22    &   0.17   \\
\hline
Side spacer      &   2.29    &   2.33   \\
\hline
Insulation film  &   66.63   &   66.67  \\
\hline
\end{tabular}
\end{table}

\begin{figure}[h!]
	\begin{center}
		\subfloat[Electric field lines in the button spacer\label{fig:field_am_a}]{%
			\includegraphics[width=0.48\textwidth]{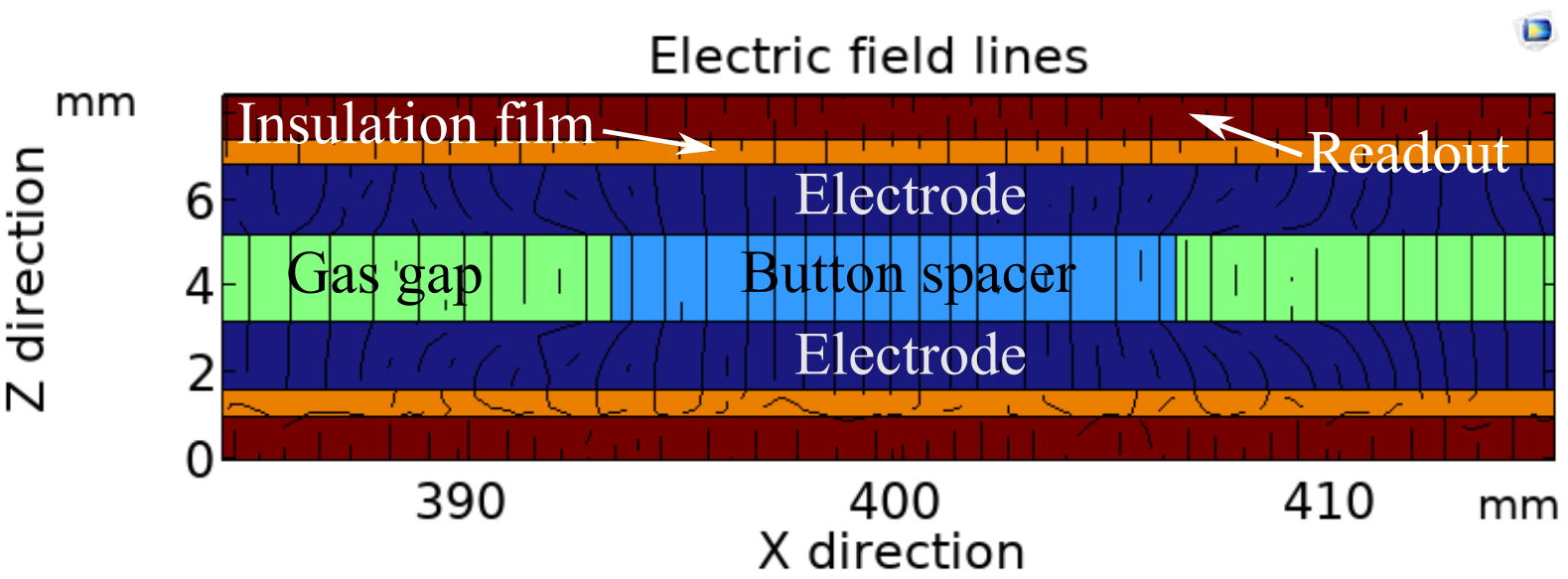}}
		\hspace{2mm}
		\subfloat[Electric field lines in the side spacer\label{fig:field_am_b}]{%
			\includegraphics[width=0.48\textwidth]{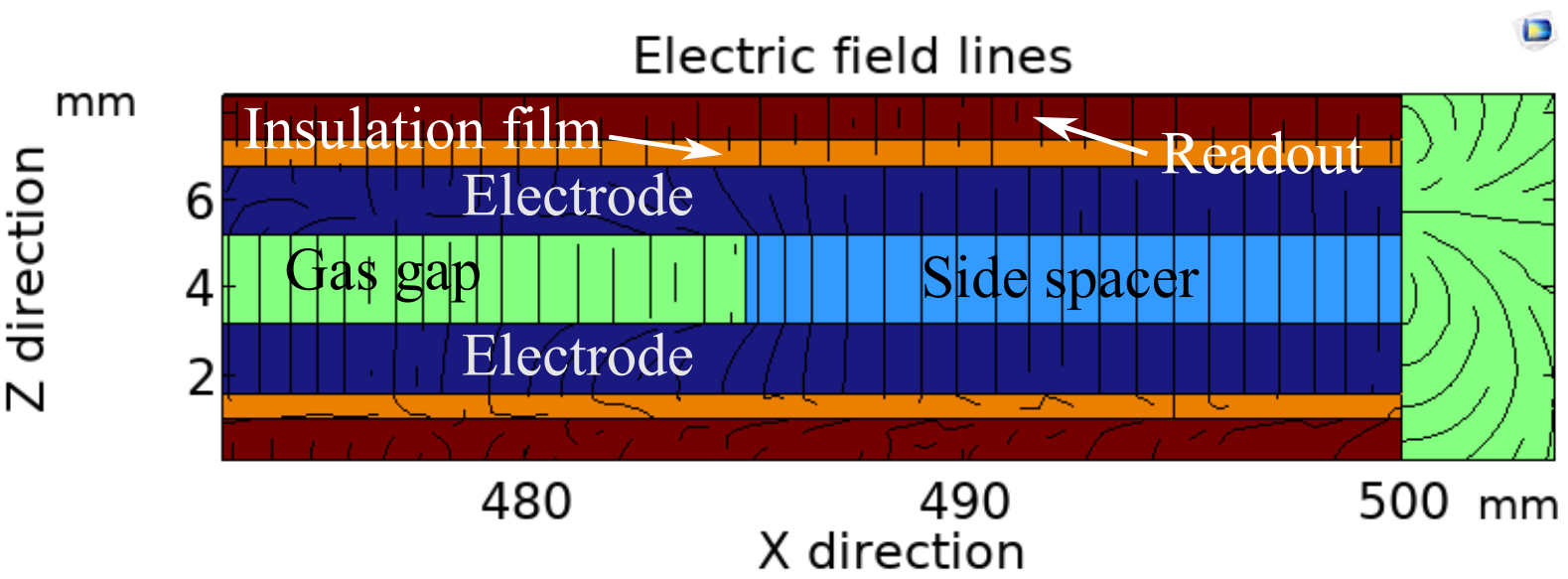}}
		\end{center}
	\caption{\label{fig:field_am} Electric field in the button and side spacers of the Bakelite RPC, modelled following \cite{Ammosov1997}}
\end{figure}

The distribution of electric field lines in the button and side spacers, as calculated by the present model, has been depicted in figure \ref{fig:field_am}.
They compare well with what was reported in \cite{Ammosov1997}. 

\subsection{\label{procon}Computational comparison between FEM and SCM}
The SCM proposed in \cite{Ammosov1997} computes the electric field as a superposition of fields arising out of surface charges and currents present on the surface and volume of a given device.
The law of current conservation is satisfied while formulating the approach.
As mentioned in \cite{Ammosov1997}, the procedure is iterative due to the inclusion of surface currents.
The iteration process converges quickly, as noted in the same reference.
Like other similar formulations based on the Green function, the approach is flexible, powerful and precise.
Due to the generation of dense matrices, the computational requirements increase rapidly with the problem size (square of the number of unknowns) despite the fact that it is necessary to discretize only the surface of a given geometry (reducing the dimension of the problem by one).
Implementation of the Fast Multipole Method (FMM) leads to significant compression and can alleviate the matrix-related problem to a good extent \cite{GREENGARD1987325, LIU2006371}.
Once solved, charge densities are available for very accurate post-processing.
The Green function based approaches, although inexpensive (often free), usually need to be developed and maintained by local groups.

The FEM when applied to electrostatic problems, on the other hand, solves for the voltage distribution throughout a given geometry and, as a result, has to discretize the entire computational domain.
Subsequent analysis of the voltage distribution leads to field values and other necessary parameters, which necessarily involve some amount of numerical errors, however small.
The flexibility of the approach (it does not depend on the existence of Green function, or similar other fundamental solutions), involvement of very sparse matrices, have led to immense popularity of the approach. 
As a result, there exists numerous extremely successful commercial products \cite{ansys,comsol,cst} that are very capable, have excellent user interfaces, geometry modelers, analysis and report generation features, in addition to very useful user support, which makes them very user-friendly but expensive.

The computational efficiency of these two approaches is comparable, although for complex, large problems the FEM is likely to be more efficient.
For the present set of problems, the time taken for FEM meshing is negligible, while the solution takes seven minutes to achieve lower than 0.01 relative error.
Here, the relative error is defined as the calculated error in the solution attempt.
The static part of the solution takes around seven minutes.
Since we do not have access to the SCM code mentioned in \cite{Ammosov1997}, a comparable neBEM \cite{MUKHOPADHYAY2009105} solver has been used for solving the static part of the problem.
The time taken by the neBEM solver is 20 minutes using OpenMP parallelization.
All the simulations have been done using a four core 2.2 GHz machine equipped with 32 GB RAM.  

\section{\label{eval} Evaluation of Electrical Characteristics of RPC}

Following its validation, the numerical model has been implemented for a systematic evaluation of the electrical characteristics of RPC for variation in electrical and geometrical parameters of the device components. 
To validate these systematic evaluations, the simulated results have also been compared to the experimental measurements done with a glass RPC.
The design specifications of the glass RPC have been described in the subsection \ref{desspec} which have been used in the numerical work to model the same.
The following observables have been evaluated and discussed in the next subsections \ref{potdis} to \ref{efield}, respectively.
\begin{itemize}
\item{Potential distribution}
\item{Electric field distribution} 
\item{Time constant of potential buildup}
\end{itemize}

\subsection{\label{desspec} Design Specifications}

\begin{figure}[h!]
    \centering
   	\includegraphics[width=0.45\textwidth]{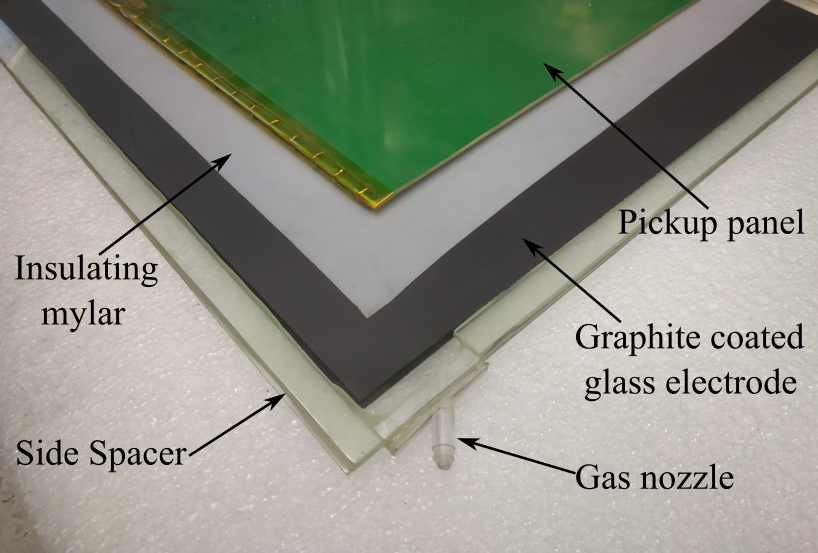}
   	\caption{Image of the fabricated glass RPC}
    \label{fig:glass_RPC}
\end{figure}

The glass RPC has been fabricated using two float glass electrodes. 
The outer surface of the electrodes have been coated with a fine layer of graphite.
The high voltages of opposite polarity have been supplied to the electrodes by connecting the leads to a small copper tape of 0.5 mm thickness pasted at one corner.
The side and button spacers used to maintain the gas gap have been made from polycarbonate (PC) which is a thermoplastic polymer.
An insulating layer of mylar or polyethylene terephthalate (PET) film has been used to isolate the graphite layer from the pickup panel.
The pickup panel has been made of copper strips pasted on a dielectric board of G10, which is a fiberglass laminate.
An image of the fabricated glass RPC has been depicted in figure \ref{fig:glass_RPC} with all its components labelled.

The glass RPC in the numerical work has been modelled following the design parameters of the fabricated one, mentioned in table \ref{table3}.
It should be noted that these properties may vary with material grade and the environment.
The actual value of these parameters for the present glass RPC is not exactly known or measured.
The electrical parameters of the glass electrode are average values that have been taken from the measurements of the same, reported in \cite{Naimuddin2014, Raveendra2016}.
For other components, some typical values of these parameters have been considered.
Due to technical limitations, the insulating mylar layer and the readout planes were not included while carrying out the experiments.
To be consistent with that, the insulating film and the readout planes beyond the conductive coat have not been included in the numerical model also.
However, these components are an important part of the electrical model to describe the entire RPC as we will illustrate in section \ref{potdis}.
\begin{table}[h!]
\centering
\caption{\label{table3}Design specifications of glass RPC components}
\begin{tabular}{ | c | c | c | c | c |  }
\hline
Component &  Dimension & Resistivity & Relative Permittivity \\
\hline
Gas gap (Gas) & $t = 2$ mm & $\rho = 10^{18} ~\SI{}{\ohm cm} $ & $\kappa$ = 1\\
\hline
Electrode  & $t = 3$ mm  & $\rho = 10^{12} ~\SI{}{\ohm cm} $ &  $\kappa$ = 10 \\
(Float glass) & $a = 300$ mm$~\times~300$ mm & &\\
\hline
Side spacer (PC) &  $t = 2$ mm, $w = 10$ mm & $\rho = 10^{14}$~\SI{}{\ohm cm} & $\kappa$ = 3 \\
\hline
Button spacer (PC) &  $t = 2$ mm, $d = 10$ mm &  $\rho = 10^{14}$~\SI{}{\ohm cm} & $\kappa$ = 3 \\
\hline
Coat (Graphite) & $t = 0.1$ mm & $R_s = 300~k\SI{}{\ohm \per \sq}$ & $\kappa$ = 10 \\
& $a = 280$ mm $~\times~280$ mm &&\\
\hline
\end{tabular}
\end{table}

\subsection{\label{potdis} Potential Distribution}

\begin{figure}[h!]
   	\begin{center}
		\subfloat[Measured $R_s$-map (top) \label{fig:Res_map_a}]{%
			\includegraphics[width=0.45\textwidth]{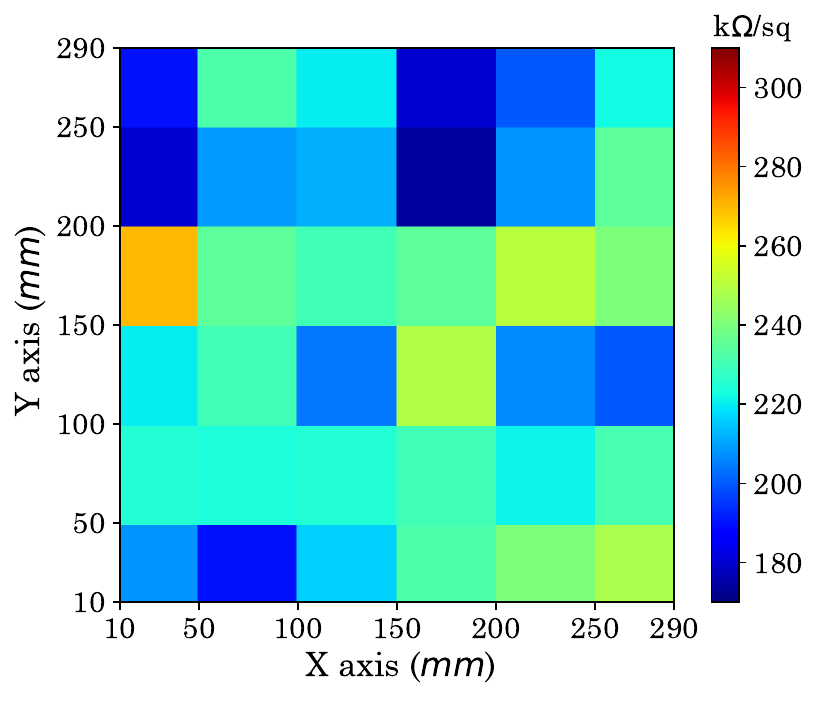}}
		\hspace{2mm}
		\subfloat[Measured $R_s$-map (bottom) \label{fig:Res_map_b}]{%
			\includegraphics[width=0.45\textwidth]{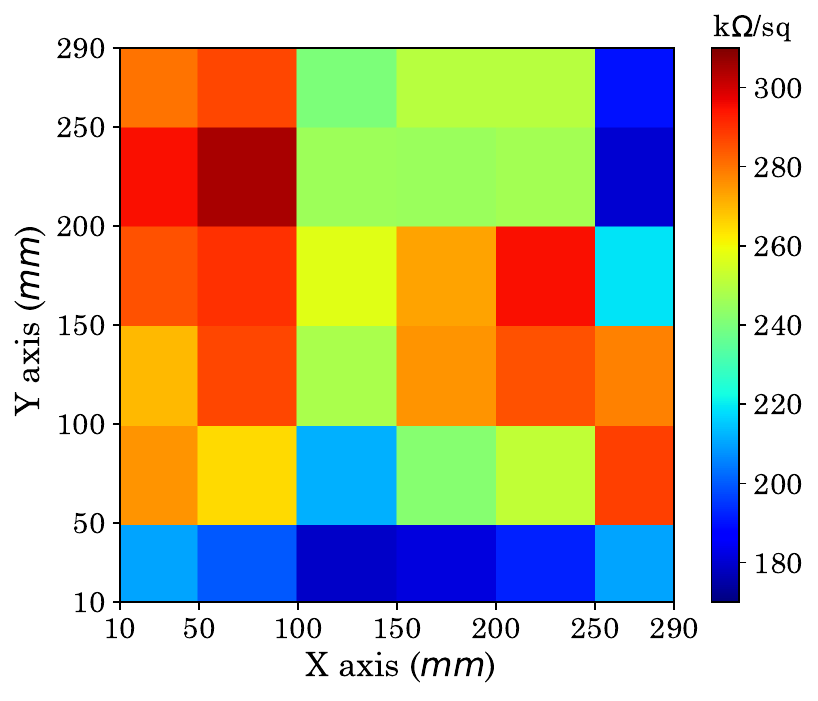}}
	\end{center}
	\caption{\label{fig:Res_map} Measured surface resistivity maps of two glass electrodes}
\end{figure}

As mentioned earlier, a high potential difference is generated between the resistive electrodes using the conductive coat applied on the electrode surface for supply of voltages.
In practice, the surface resistivity of the glass electrode coated with graphite depends on the thickness of the coat, which may often vary depending upon the procedure.
The surface resistivity of the glass electrode coated with graphite by spray-painting method has been measured in the following way.
The surface resistance of a square region of area $5~cm \times 5~cm$ of the coated electrode has been determined by measuring the current flowing through a surface resistivity measurement probe using picoammeter (\textit{model: KEITHLEY 6487}) when a constant potential has been applied across it.
A map of surface resistivity ($R_s$-map) of the entire electrode thus has been produced from the measured data of 36 square cells.
Due to the dimension mismatch between the coated area (280 mm $\times$ 280 mm) and the square shaped probe (50 mm $\times$ 50 mm), while measuring the surface resistivity, there is an overlap between the last two rows of cells.
While carrying out the measurement, it was kept in mind that the probe is only on the graphite coating.
The same for each of the top and bottom glass electrodes respectively has been displayed in figures \ref{fig:Res_map_a} and \ref{fig:Res_map_b}.
The plots show that the surface resistivity values are distributed over a range of $180 - 320~ k\SI{}{\ohm\per \sq}$ on both the electrodes.
 
\begin{figure}[h!]
   	\begin{center}
		\subfloat[Components and experimental setup \label{fig:Pot_meas_a}]{%
			\includegraphics[width=0.50\textwidth]{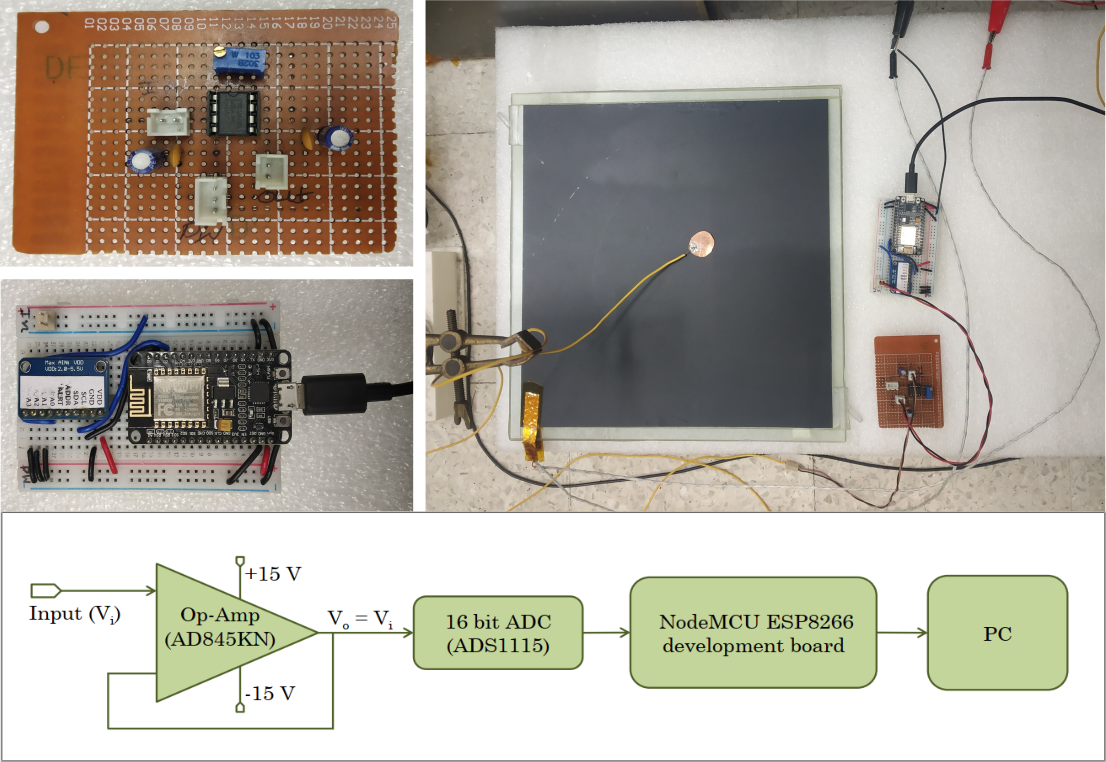}}
		\hspace{2.5mm}
		\subfloat[Distribution of measured potential at a location \label{fig:Pot_meas_b}]{%
			\includegraphics[width=0.445\textwidth]{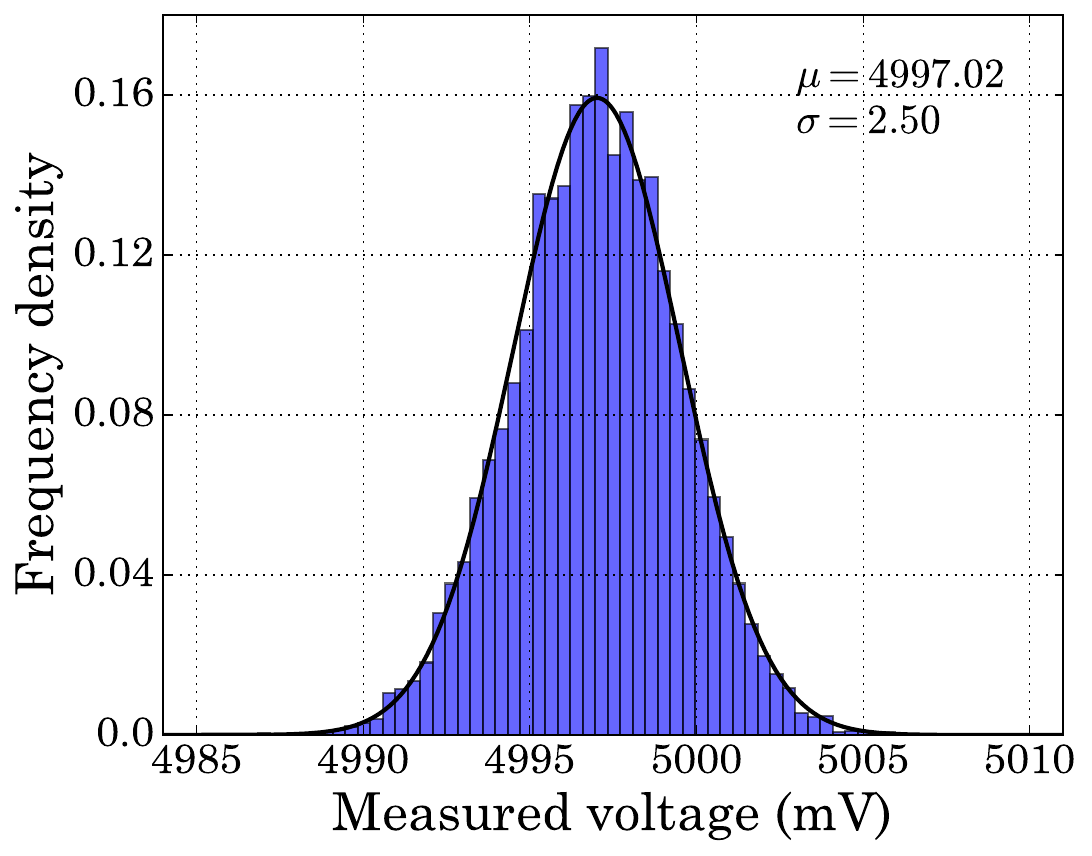}}
	\end{center}
	\caption{\label{fig:Pot_meas} Potential measurement on top electrode of glass RPC}
\end{figure}

\begin{figure}[h]
   	\begin{center}
		\subfloat[Measured potential map\label{fig:Pot_comp_a}]{%
			\includegraphics[width=0.48\textwidth]{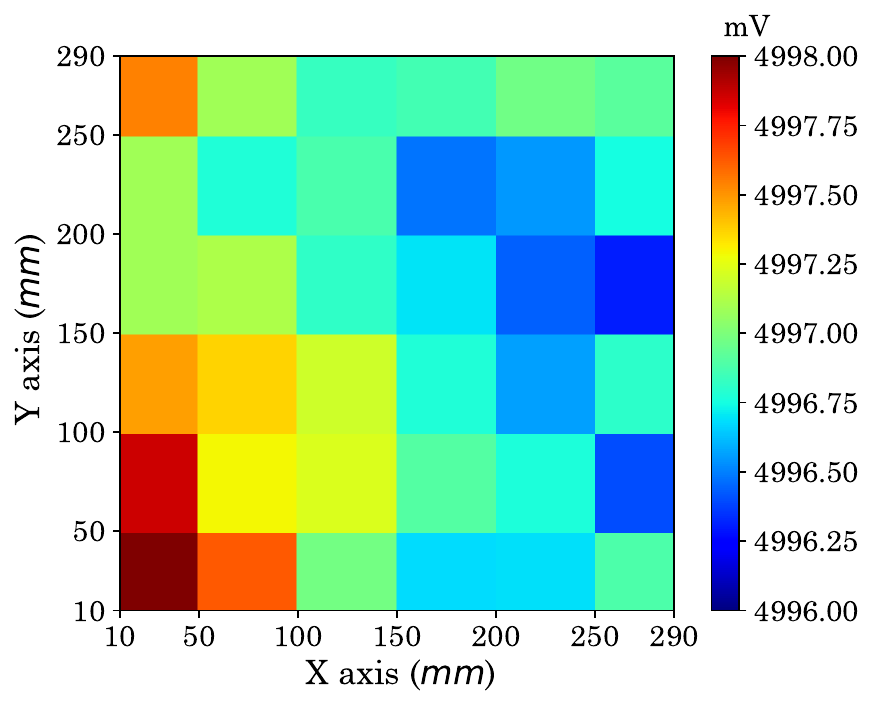}}
		\hspace{2 mm}
		\subfloat[Simulated potential distribution\label{fig:Pot_comp_b}]{%
			\includegraphics[width=0.48\textwidth]{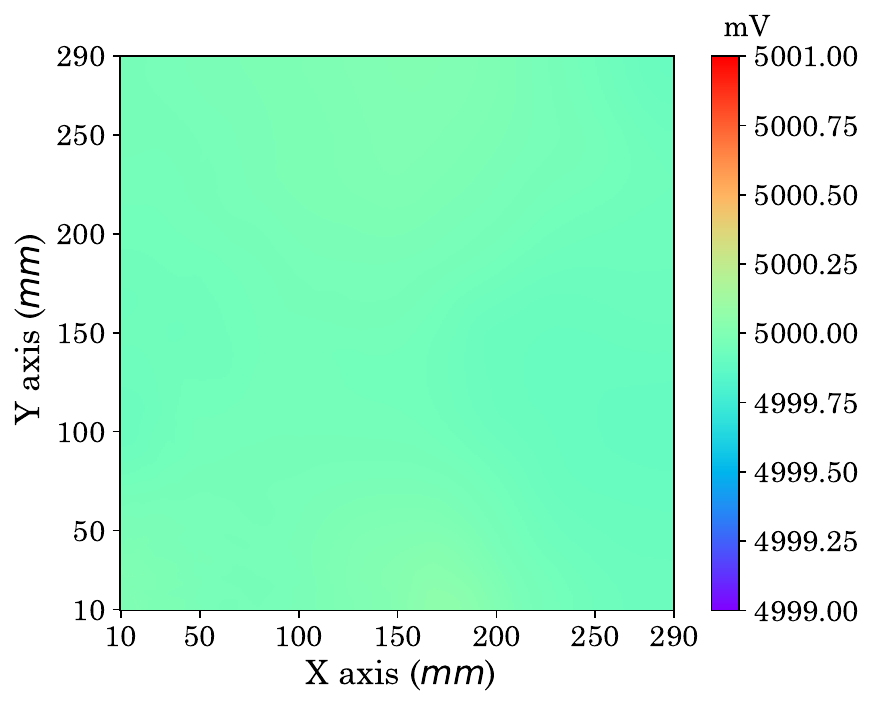}}
	\end{center}
	\caption{\label{fig:Pot_comp} Measured and simulated potential distribution on top electrode of glass RPC}
\end{figure}

In order to study the effect of the non-uniform surface resistivity, the potential distribution on the glass electrode (top) has been measured for a given supply of $5~V$.
A DC power supply (\textit{model: Agilent E3631A}) with a precision of $1~mV$ has been used for this purpose. 
The order of input impedance of a multimeter ($10~M\SI{}{\ohm}$) used for measuring the potential is comparable to the surface resistivity of graphite coat ($180 - 320~k\SI{}{\ohm\per \sq}$). 
This can cause a drop in the measured potential.
In order to circumvent the issue, a voltage follower circuit (buffer circuit) has been designed using an Op-Amp (\textit{model: AD845KN}) with a very high input impedance ($10^{11}~k\SI{}{\ohm}$) and a low output impedance.
A 16 bit ADC (\textit{model: ADS1115}) has been used for precise measurements.
The complete setup using the buffer circuit, ADC, and a development board (\textit{model: ESP8266}) with Tensilica Xtensa 32 bit LX106 RISC microprocessor, has been able to measure 0 to $6.144~V$ with the least count of $0.1875~mV$.
The components of the experimental setup and its arrangement along with a schematic diagram of the same have been depicted in figure \ref{fig:Pot_meas_a}.
Due to RF noise and limited accuracy of the power supply and the experimental setup, some fluctuations have been observed while measuring the potential.
The histogram of 1000 measurements done at a particular location has been  shown in figure \ref{fig:Pot_meas_b}.
A map of the measured potential values of the glass electrode (top) has been produced from the mean of the histogram obtained at different locations.
The potential map has been displayed in figure \ref{fig:Pot_comp_a} which has a very little fluctuation arising due to the experimental limitations of the measurement.
Therefore, it is implied that for a fluctuation in the surface resistivity over the range of $180~-~320~k\SI{}{\ohm\per \sq}$, the potential gets uniformly distributed.

The potential distribution on the glass electrode  has been numerically evaluated using the measured $R_s$-map of the same (top), as shown in figure \ref{fig:Res_map_a}.
The result is shown in figure \ref{fig:Pot_comp_b} which displays a uniform potential distribution.
This demonstrates a nice agreement of the numerical model with the experimental observation.

\begin{figure}[h]
   	\begin{center}
   	    \subfloat[Simulated potential without mylar and readout planes.\label{fig:Potres_rho_elec_worp}]{%
			\includegraphics[width=0.48\textwidth]{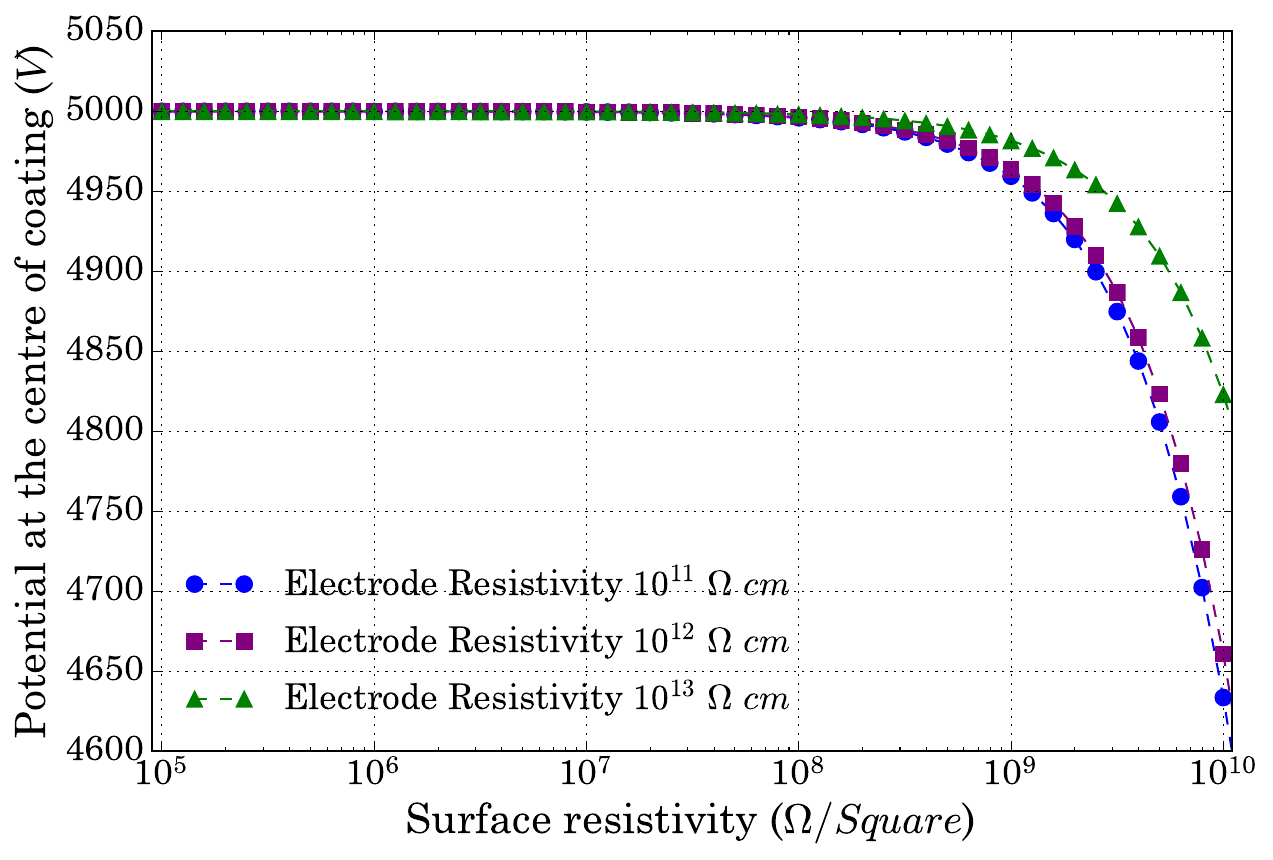}}
		\hspace{2 mm}
		\subfloat[Simulated potential with mylar and readout planes. \label{fig:Potres_rho_elec_wrp}]{%
			\includegraphics[width=0.48\textwidth]{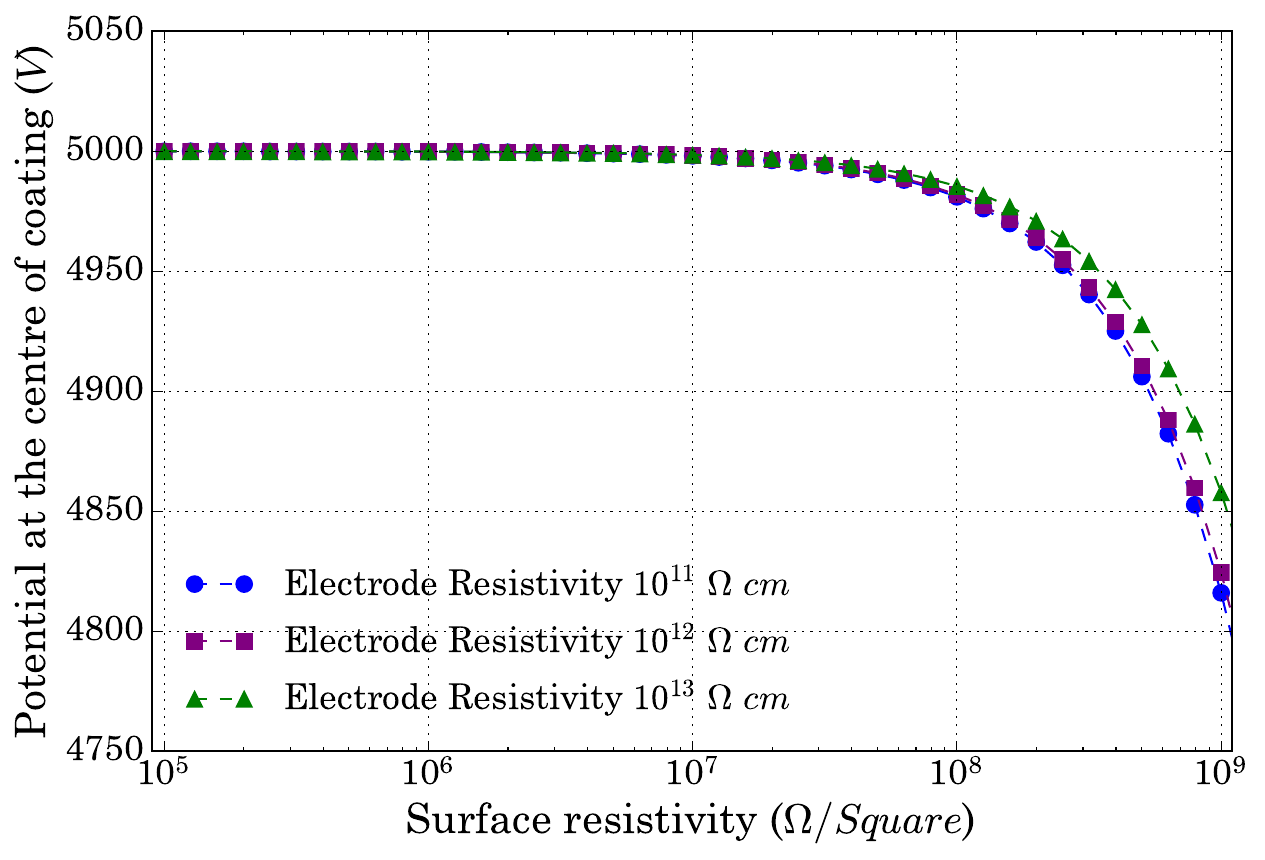}}
	\end{center}
	\caption{\label{fig:Potres_rho_elec} Simulated potential at the center of the conductive coating as a function of surface resistivity and for different volume resistivity of electrode}
\end{figure}

\begin{figure}[h]
    \centering
    \includegraphics[width=0.5\textwidth]{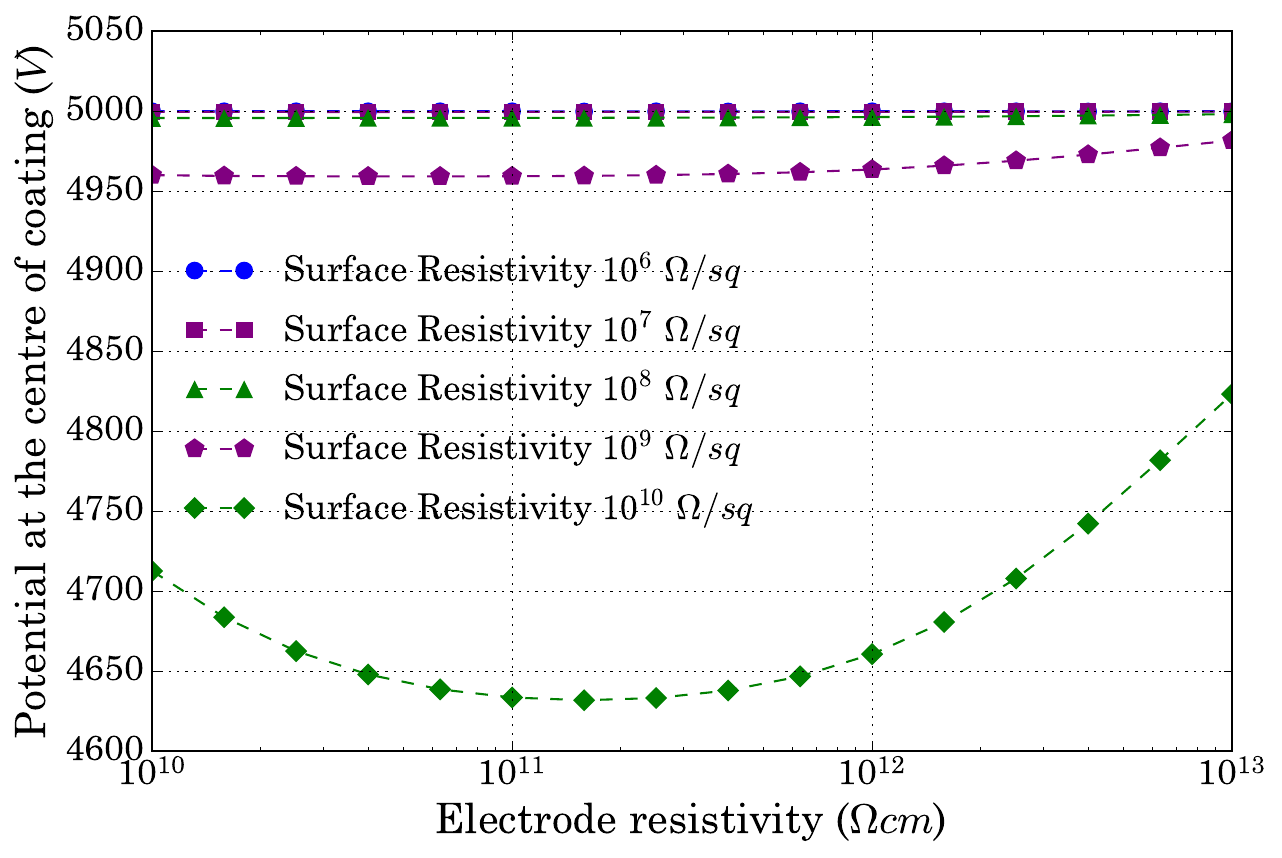}
    \caption{Simulated potential at the center of the conductive coating as a function of the electrode volume resistivity for different surface resistivity of the conductive coating}
    \label{fig:var_elec_resis}
\end{figure}

To study the effect of the surface resistivity on the potential in general, a numerical evaluation has been performed to determine the potential at the center of the electrode as a function of surface resistivity.
The electrodes are assumed to be at $\pm$ 5000 V respectively, making the potential drop across the gas gap 10 kV.
The calculation has been repeated for a change in volume resistivity of the electrode over $\rho = 10^{11}$ to $10^{13}~\SI{}{\ohm cm}$ which includes the typical materials (glass, Bakelite) for making the electrodes.
The other components of the RPC have been kept the same as described in the table \ref{table3}.
The results have been plotted in figure \ref{fig:Potres_rho_elec}.
The simulated results without the insulating layer and readout planes, shown in figure \ref{fig:Potres_rho_elec_worp}, depicts that the potential at the center of the electrode can be held uniform for a wide range of variation in the surface resistivity ($R_s = 100~k\SI{}{\ohm\per \sq} - 1~G\SI{}{\ohm\per \sq}$).
However, when the simulation is carried out considering the mylar and the readout planes, the potential at the center of the electrode can be held uniform for a smaller range of the surface resistivity ($R_s = 100~k\SI{}{\ohm\per \sq} - 100~M\SI{}{\ohm\per \sq}$), shown in figure \ref{fig:Potres_rho_elec_wrp}.
In absence of the readout planes and mylar, the potential is least affected by change in the electrode material until the surface resistivity becomes as high as $R_s =1~G\SI{}{\ohm\per \sq}$ or more.
As the volume resistivity of the electrode is substantially high in comparison to its surface resistivity, it offers high resistance for the dark current to flow through it.
Therefore, the maximum amount of current flows through the surface and hence, a change in the volume resistivity does not affect the situation.
However, the volume resistivity starts to affect when the surface resistivity becomes too high to allow the current flowing through the surface.
Then, a substantial amount of current flows through the electrode and the change in its volume resistivity becomes a contributing factor.
This phenomenon is evident from the figure \ref{fig:var_elec_resis}, where the simulated potential has been shown as a function of the electrode's volume resistivity for different surface resistivity of the conductive coating.
In the presence of the mylar, an additional path becomes available for current flow, which justifies the larger drop in potential in presence of mylar and the readout planes.

\begin{figure}[h]
	\begin{center}
		\subfloat[$R_s$ = 1 M\SI{}{\ohm\per \sq}\label{fig:Pot_rs_a}]{%
			\includegraphics[width=0.32\textwidth]{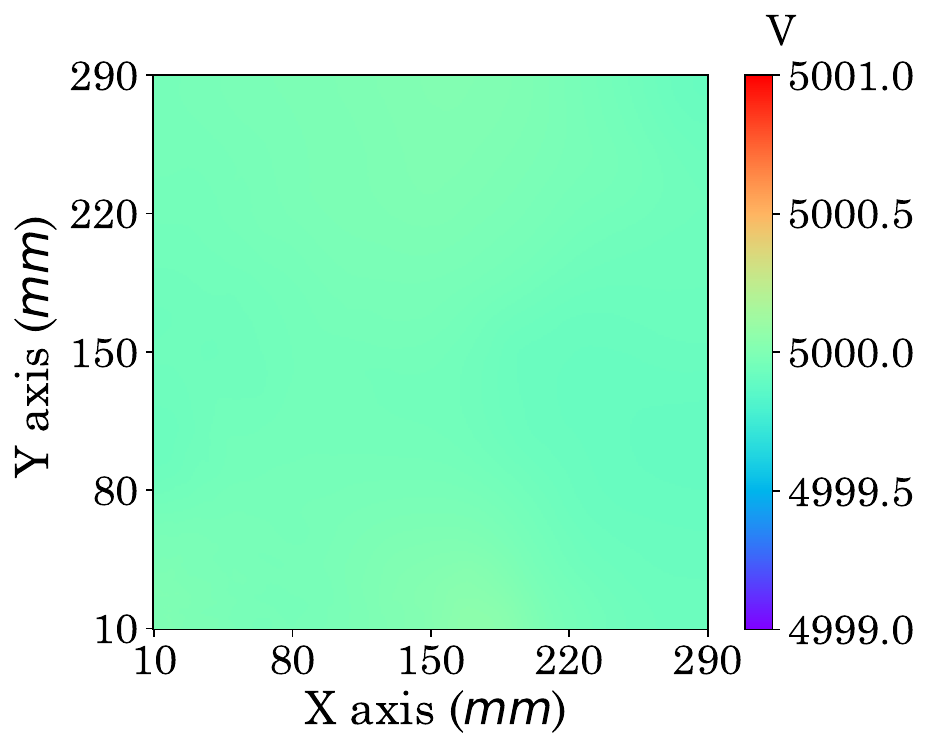}} 
		\hspace{2mm}
		\subfloat[$R_s$ = 100 M\SI{}{\ohm\per \sq}\label{fig:Pot_rs_b}]{%
			\includegraphics[width=0.32\textwidth]{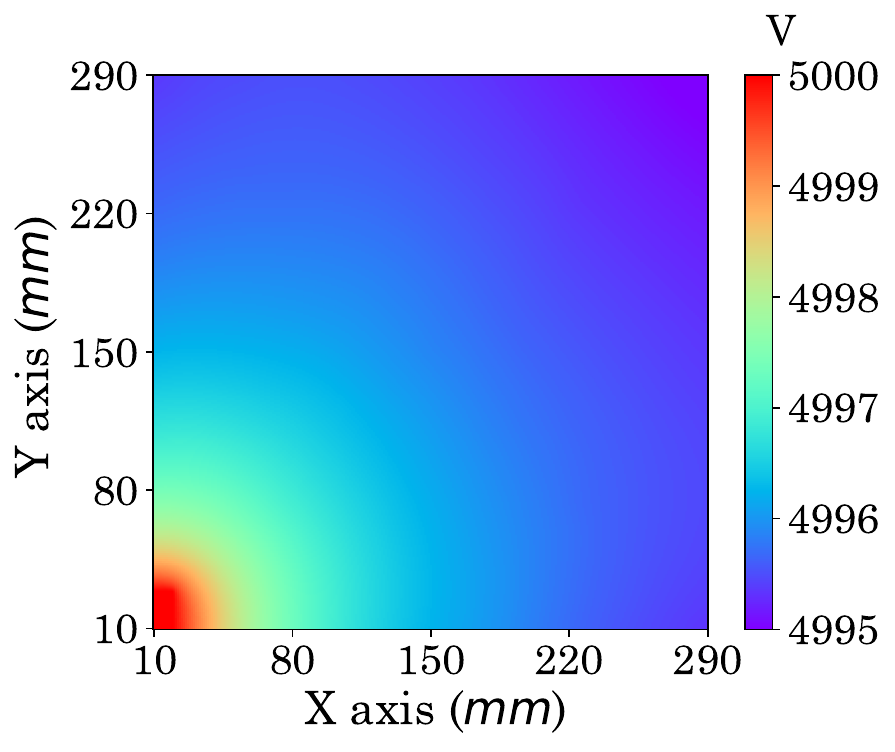}}
		\hspace{2mm}
		\subfloat[$R_s$ = 1 G\SI{}{\ohm\per \sq}\label{fig:Pot_rs_c}]{%
			\includegraphics[width=0.32\textwidth]{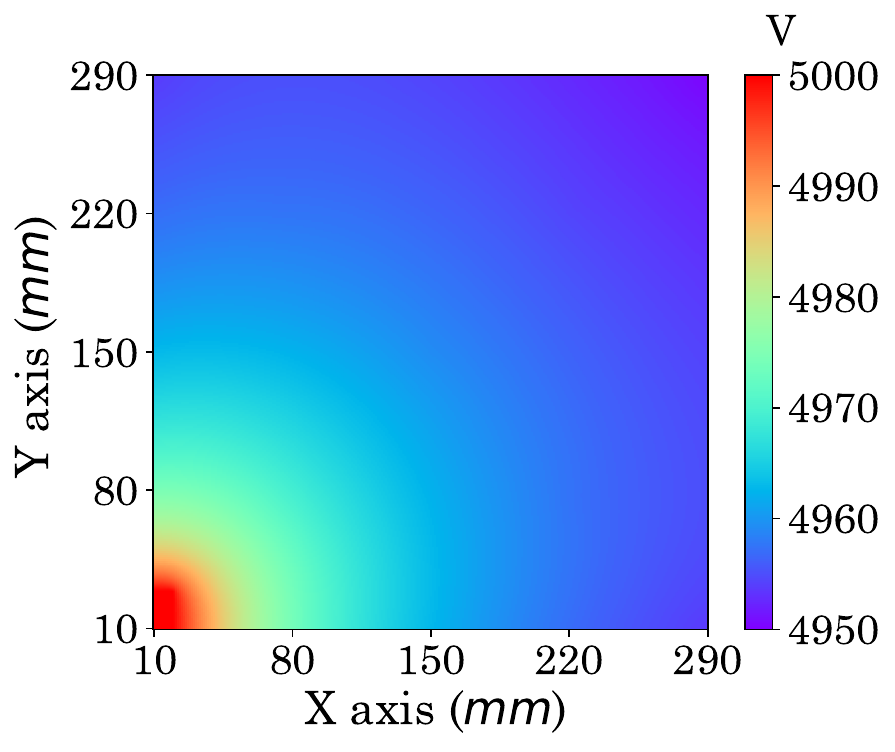}}
	\end{center}
	\caption{\label{fig:Pot_rs}Potential distribution on glass electrode at three different values of surface resistivity. The HV is applied at the left bottom corner of the conductive coating}
\end{figure}

The potential distribution over the entire coated area of the glass electrode of dimension $280 \times 280~mm^2$ for actual high voltage supply of $\pm 5000~V$  has been studied as well using the numerical model.
It has been computed for three different values of the surface resistivity of the graphite coat ($R_s = 1~M\SI{}{\ohm\per \sq}, 100~M\SI{}{\ohm\per \sq}, 1~G\SI{}{\ohm\per \sq}$).
The results have been illustrated in figure \ref{fig:Pot_rs}.
It shows that a nominal drop in the potential ($\sim$ 1\%) is observed across the electrode at substantially high value of surface resistivity of $R_s=1~G\SI{}{\ohm\per \sq}$.
For larger dimension of the electrodes, a bigger drop in the potential is implied.

\begin{figure}[h!]
	\begin{center}
		\subfloat[Simulated potential without mylar and readout planes.\label{fig:Potres_rho_spac_worp}]{%
			\includegraphics[width=0.48\textwidth]{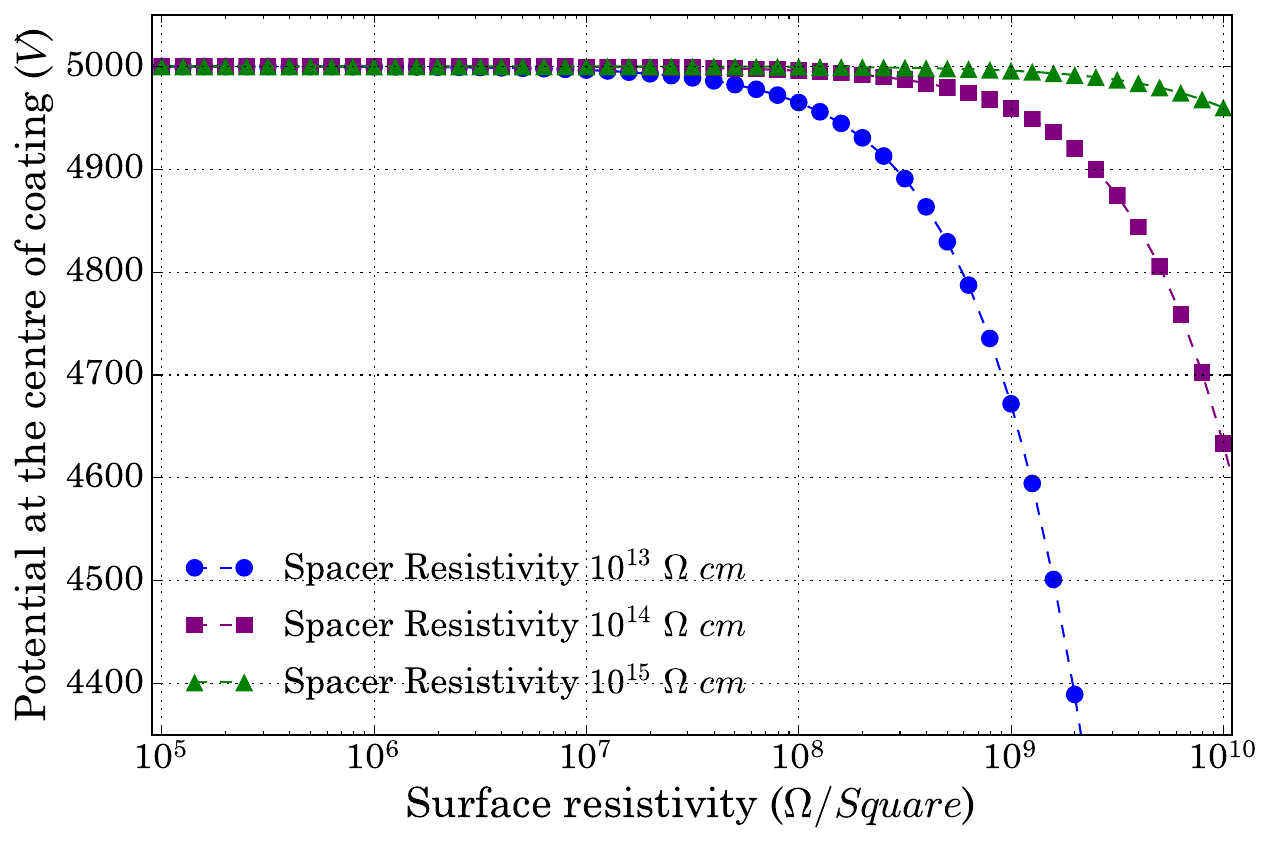}}
		\hspace{2mm}
		\subfloat[Simulated potential with mylar and readout planes.\label{fig:Potres_rho_spac_wrp}]{%
			\includegraphics[width=0.48\textwidth]{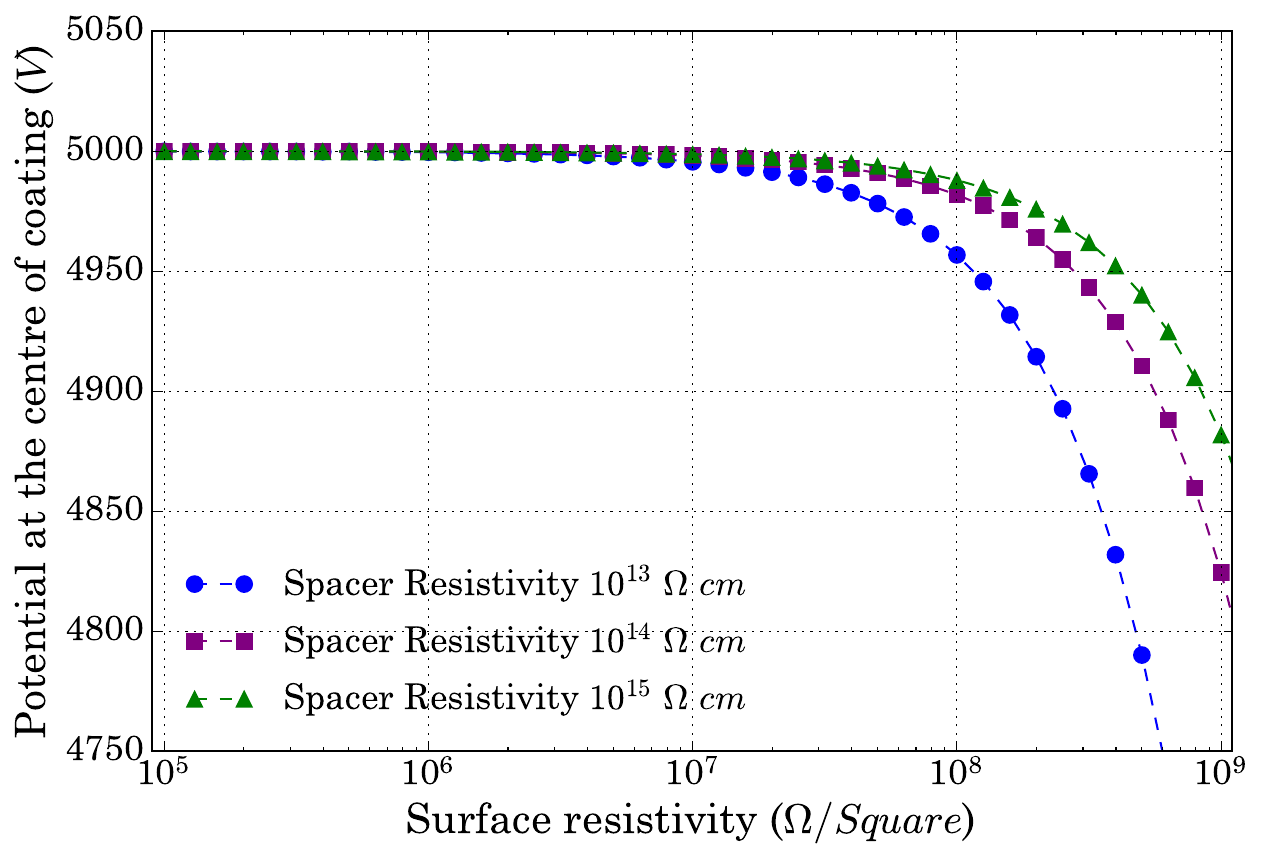}}
		\end{center}
	\caption{\label{fig:Potres_rho_spac}Potential at the center as a function of surface resistivity and for different values of volume resistivity of spacers. The volume resistivity of the electrodes has been kept fixed at $10^{12} ~\SI{}{\ohm cm}$}.
\end{figure}

The other device components which can be instrumental in governing the potential distribution are the spacers (button and side) as they provide a path for the dark current to flow through.
A similar calculation has been performed to study the effect of the volume resistivity of the spacers on the potential distribution over the electrodes with and without considering the readout planes.
The results for with and without considering the readout planes has been displayed in figures \ref{fig:Potres_rho_spac_worp} and \ref{fig:Potres_rho_spac_wrp} respectively, which suggest that higher value of the spacer resistivity helps to sustain the uniform potential for a wider range of variation in surface resistivity of the electrode.
But the presence of the readout has a negative effect, as shown in figure \ref{fig:Potres_rho_spac_wrp}.
The additional path for dark current through the insulating layer is to be blamed for that.
From figures \ref{fig:Potres_rho_elec_wrp} and \ref{fig:Potres_rho_spac_wrp} it is evident that the presence of the mylar and the readout planes have a significant effect only when the surface resistivity of the conductive coating  becomes as large as $100~M\SI{}{\ohm\per \sq}$.
But all the experimental measurements carried out in this work assumes a surface resistivity of the order of $100~k\SI{}{\ohm\per \sq}$, at which the effect of the insulating layer and readout planes are negligible.
Along with the experimental limitations, this also motivates the numerical simulations without the mylar and the readout planes for the rest of the work.
It can be noted that the ratio of the volume resistivity of the spacer and the electrode materials is an important parameter to have a uniform potential throughout the conductive coating.
A value of 10 of this ratio can allow a surface resistivity of, $R_s = 100~M\SI{}{\ohm\per \sq}$ whereas when this ratio is $10^3$ it allows the surface resistivity up to $R_s = 10~G\SI{}{\ohm\per \sq}$.

\begin{figure}[h!]
	\begin{center}
		\subfloat[Current density through button spacer\label{fig:Curr_den_a}]{%
			\includegraphics[width=0.48\textwidth]{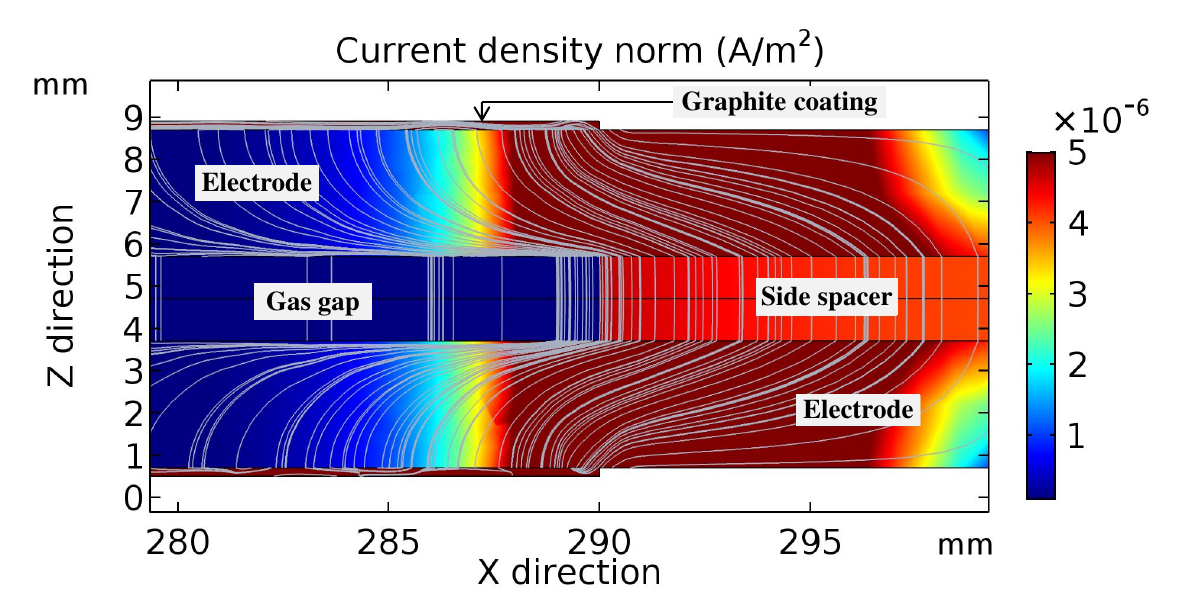}}
		\hspace{2mm}
		\subfloat[Current density through side spacer\label{fig:Curr_den_b}]{%
			\includegraphics[width=0.48\textwidth]{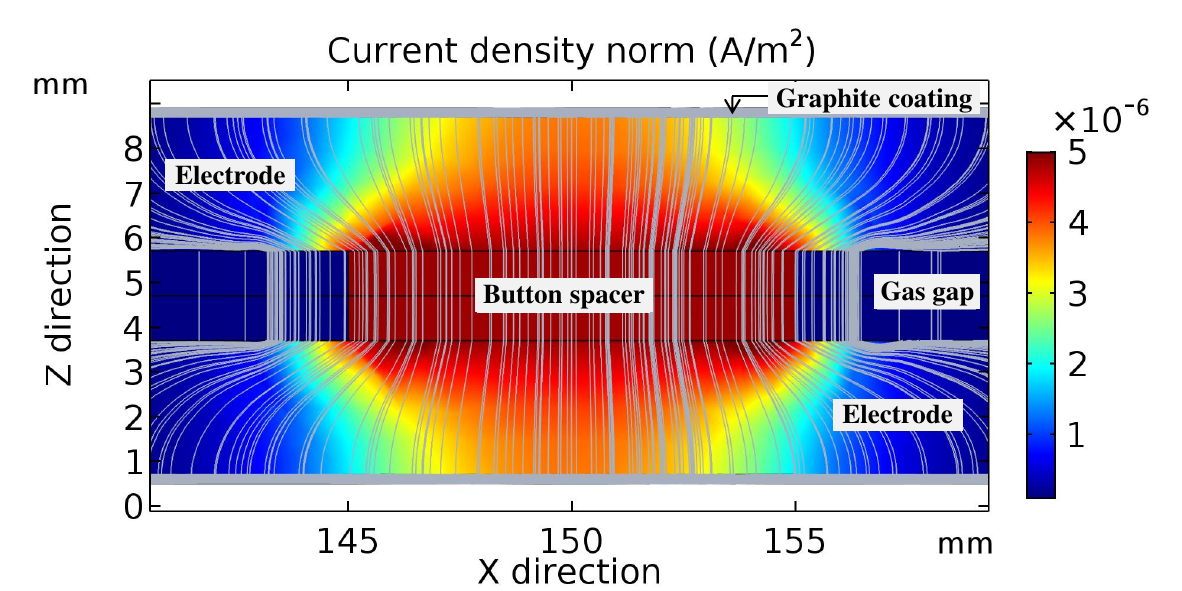}}
		\end{center}
	\caption{\label{fig:Curr_den}Current density through button and side spacers}
\end{figure}

In this context, two plots of current density in the button and side spacers, as illustrated in figures \ref{fig:Curr_den_a} and \ref{fig:Curr_den_b} respectively, may be found useful to explain the previous observations.
It is evident from the figures that most of the dark current finds its way through the spacers, as their volume resistivity is much smaller ($\rho = 10^{14}~\SI{}{\ohm cm}$) in comparison to that of the gas volume ($\rho = 10^{18}~\SI{}{\ohm cm}$).
It is obvious that a comparatively lower resistivity of the spacer will allow more dark current to pass through it, causing a drop in the potential on the electrode.
Therefore, the relative ratio of the spacer to electrode resistivity turns out as a key factor in governing the potential distribution.

\begin{figure}[h!]
    \centering
    \includegraphics[width=0.6\textwidth]{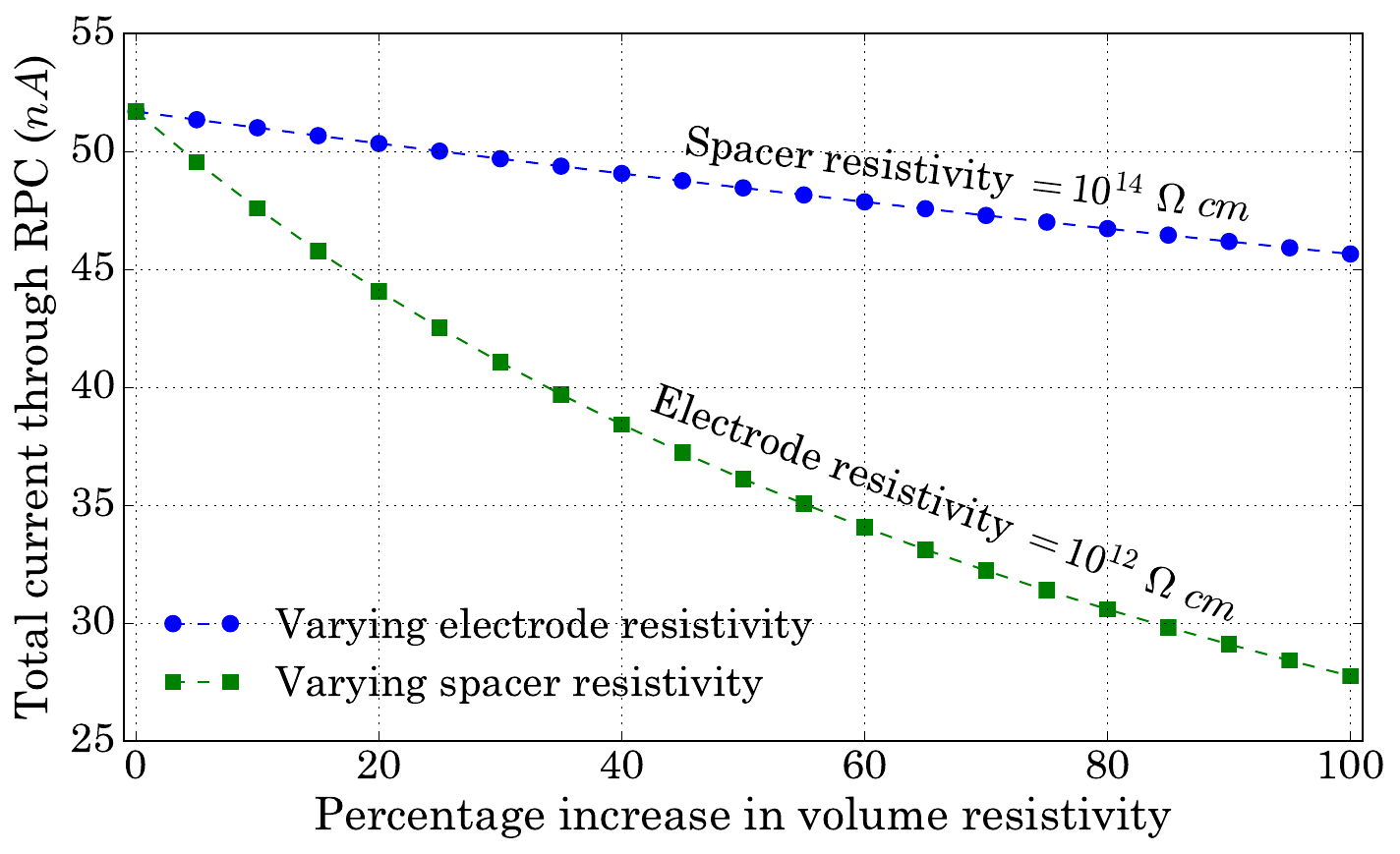}
    \caption{Dark current with percentage of variation in volume resistivity of electrode and spacer}
    \label{fig:Curr_varho}
\end{figure}

A comparison of the effect of volume resistivity of the electrodes and the spacers on the dark current is shown in figure \ref{fig:Curr_varho} for a comprehensive representation.
Here, the change in the dark current has been plotted as a function of percentage increase in their volume resistivity, respectively.
The blue line in the plot indicates the change in the dark current when the spacer resistivity has been kept fixed at $\rho = 10^{14}~\SI{}{\ohm cm}$, while increasing the electrode resistivity of $\rho = 10^{12}~\SI{}{\ohm cm}$ by $10-100\%$.
The green line in the plot indicates the reverse process, where the electrode resistivity has been kept constant at $\rho = 10^{12}~\SI{}{\ohm cm}$ and the spacer resistivity of $\rho = 10^{14}~\SI{}{\ohm cm}$ has been raised.
It is evident from figure \ref{fig:Curr_varho} that the change in spacer resistivity is more influential in governing the dark current than the electrode resistivity is.
A 60\% increase in the volume resistivity of spacer material reduces the dark current by about 30\% and also helps to hold the potential on the resistive electrode.

\subsection{\label{efield} Electric Field Distribution}

\begin{figure}[h!]
	\begin{center}
		\subfloat[Ratio of $\rho_{spacer}$  to $\rho_{electrode} = 10^3$\label{fig:field_2d_a}]{%
			\includegraphics[width=0.45\textwidth]{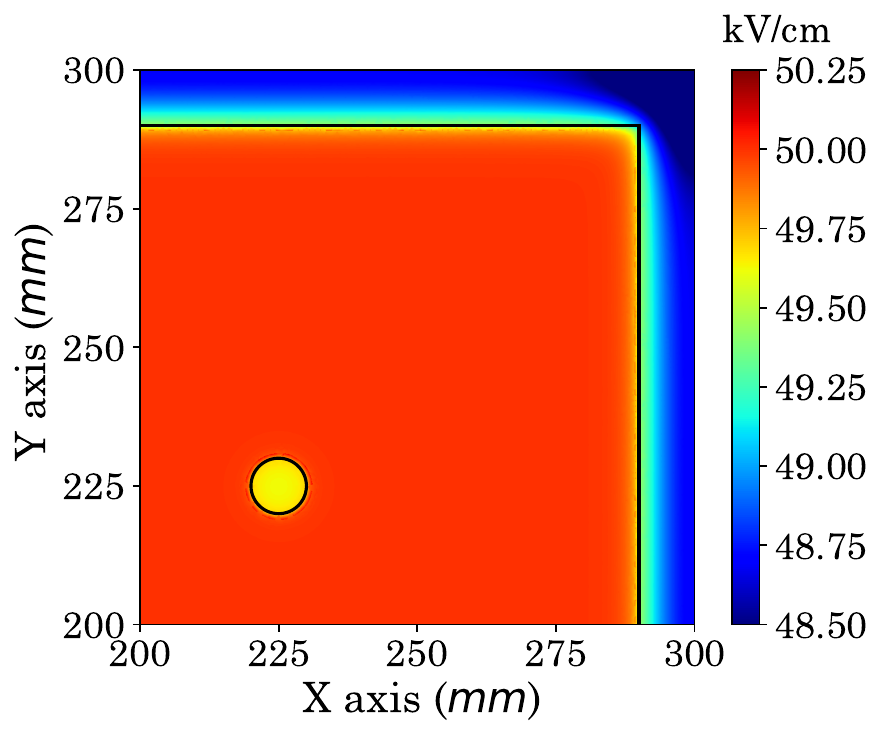}}
		\hspace{2mm}
		\subfloat[Ratio of $\rho_{spacer}$ to $\rho_{electrode} = 10$\label{fig:field_2d_b}]{%
			\includegraphics[width=0.45\textwidth]{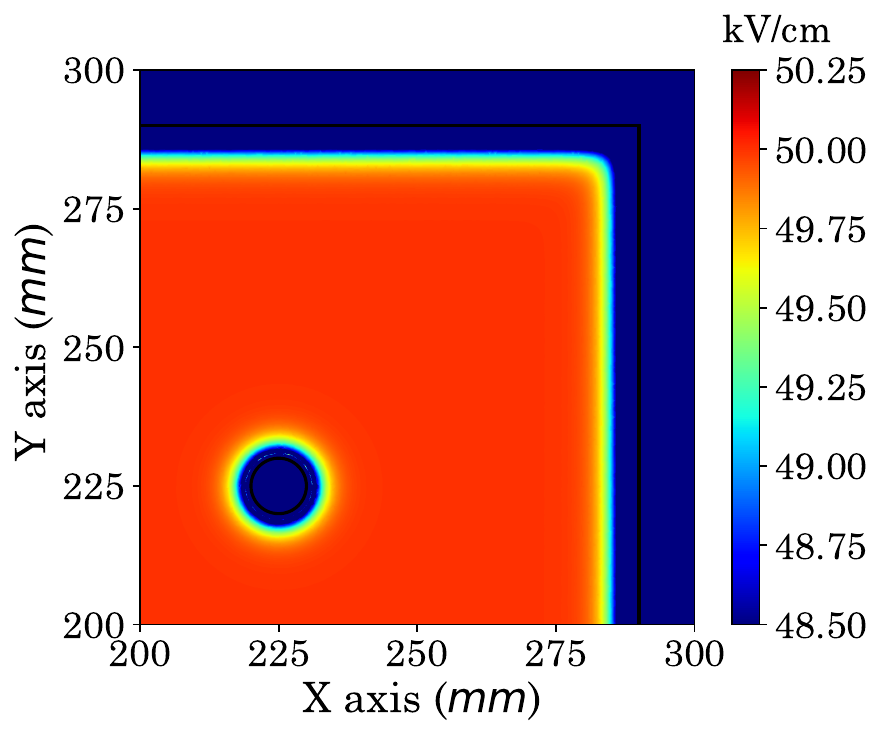}}
	\end{center}
	\caption{\label{fig:field_2d}Electric field distribution in the gas gap for two different ratios of spacer to electrode resistivity}
\end{figure}

It can be followed from the previous discussions that the volume resistivity of the electrodes and spacers play important roles in governing the electric field in the device in a combined manner.
The field maps in the gas gap for two different ratios ($10^3$ and 10) of the volume resistivity of the spacer to that of the electrode have been depicted in figures \ref{fig:field_2d_a} and \ref{fig:field_2d_b} for a supply of $\pm 5000~V$ to the glass RPC.
As most of the dark current flows through the side and button spacers, the components of current density organize themselves in such a way that the area of the electrode in contact with the gas gap realizes minimum current.
Due to this, the electrode acts as transparent material to the applied potential.
Thus, the potential applied on the conductive coat of the resistive electrode can actually be observed across the gas gap.
With the given configuration and applied voltage, an electric field of $50~kV/cm$ is expected across the gas gap, which is evident from the plots.
It can also be noted that a smaller ratio of the volume resistivity of the spacer to that of the electrode causes a larger distortion in the electric field distribution, as shown in figure \ref{fig:field_2d_b}.
The boundary of the spacers have been marked with solid lines in both the plots.
Therefore, the choice of spacer material turns out to be an important factor for consideration in designing an RPC device.

\begin{figure}[h!]
   \begin{center}
		\subfloat[Electric field near button spacer\label{fig:field_edge_a}]{
			\includegraphics[width=0.45\textwidth]{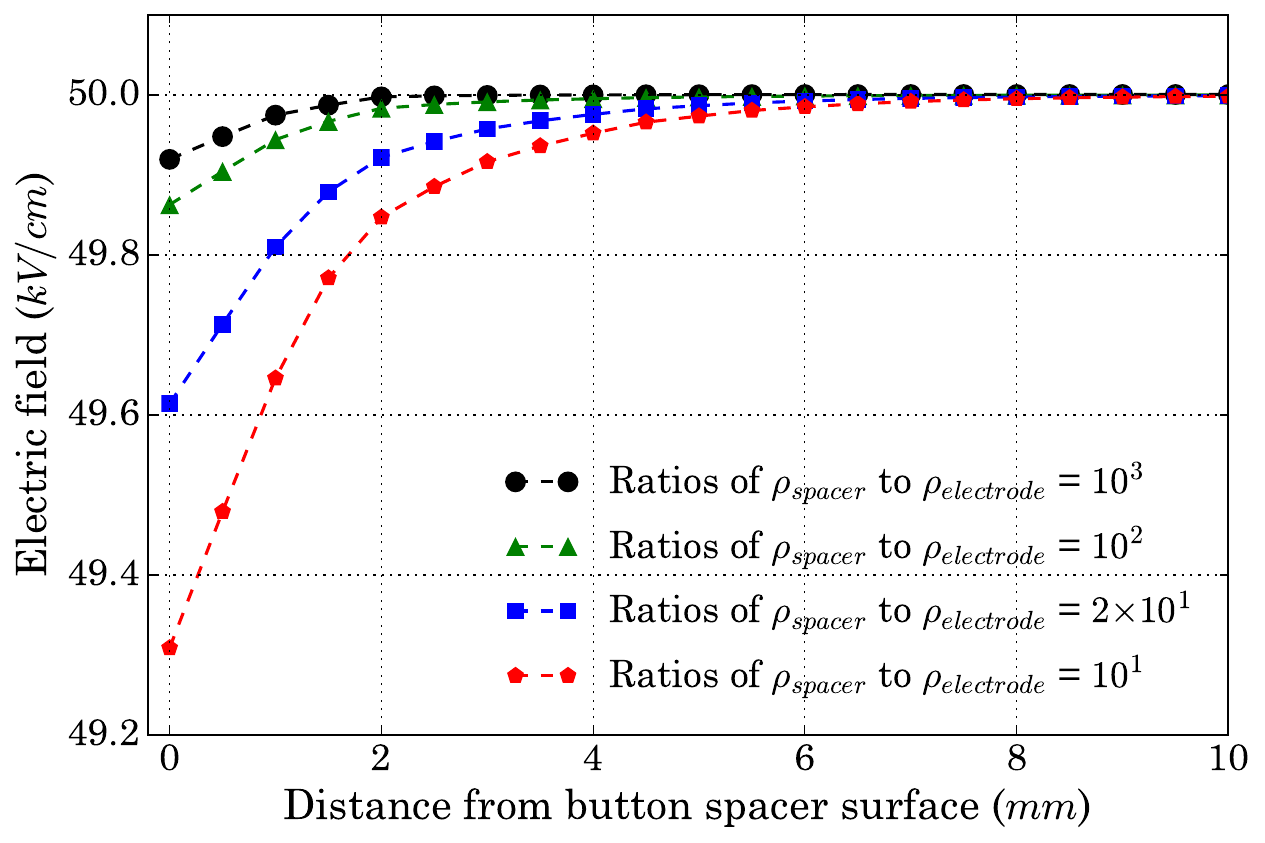}}
		\hspace{2mm}
		\subfloat[Electric field near corner of side spacers\label{fig:field_edge_b}]{
			\includegraphics[width=0.45\textwidth]{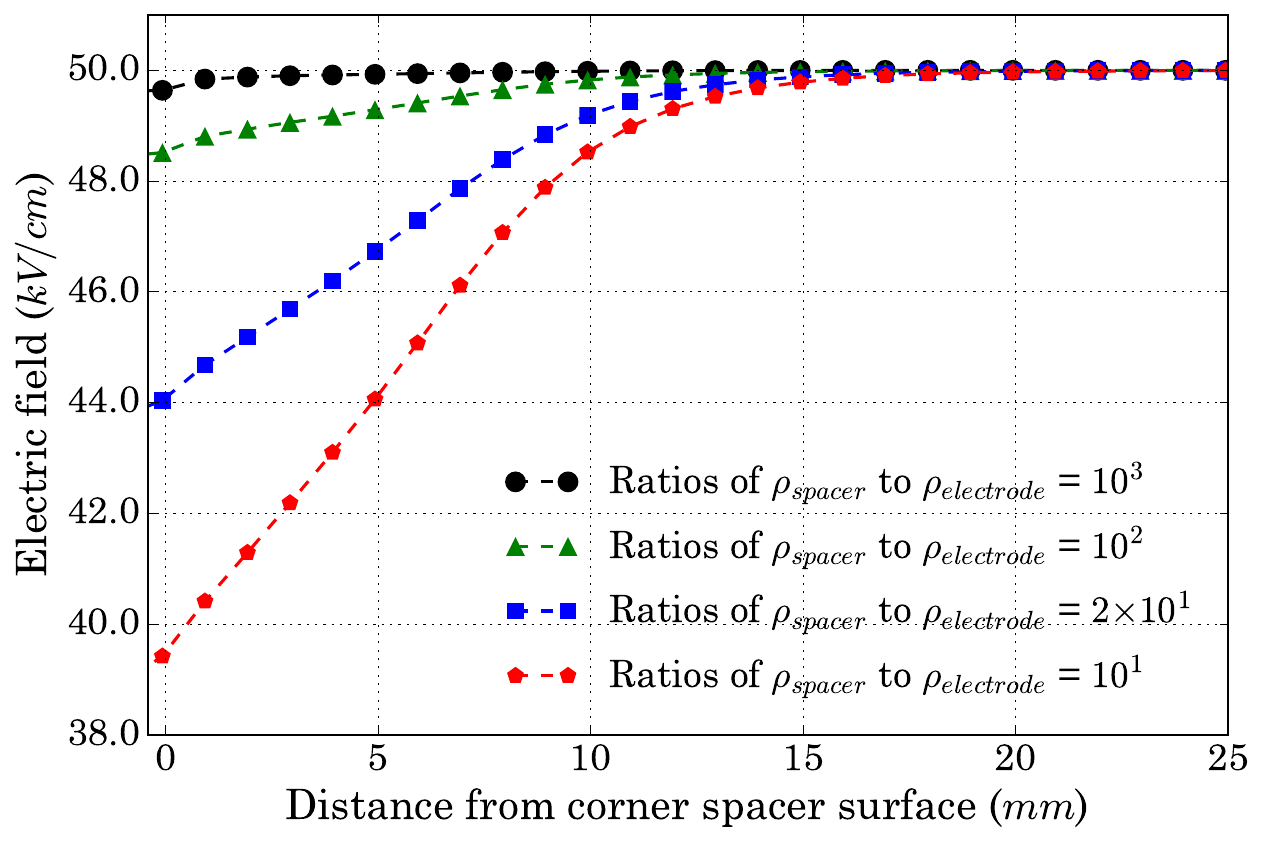}}
	\end{center}
	\caption{\label{fig:field_edge}Electric field as a function of the distance from the edge of the button spacer and corner of the side spacers for different ratios of spacer to electrode resistivity}
\end{figure}

The extent of the field distortion near the spacers is shown in figures \ref{fig:field_edge_a} and \ref{fig:field_edge_b}.
Here, the variation in the field value as the function of distance from the edge of the button and the corner of the side spacers has been plotted.
Both the plots indicate that the field gets more distorted in case of lower ratio of the spacer to electrode resistivity and reaches the uniform value at a larger distance.

\subsection{\label{tc} Time Constant of Potential Buildup}

\begin{figure}[h!]
   \begin{center}
		\subfloat[Test setup for measurement of potential buildup time constant\label{fig:Meas_timecons_a}]{
			\includegraphics[width=0.35\textwidth, angle=270]{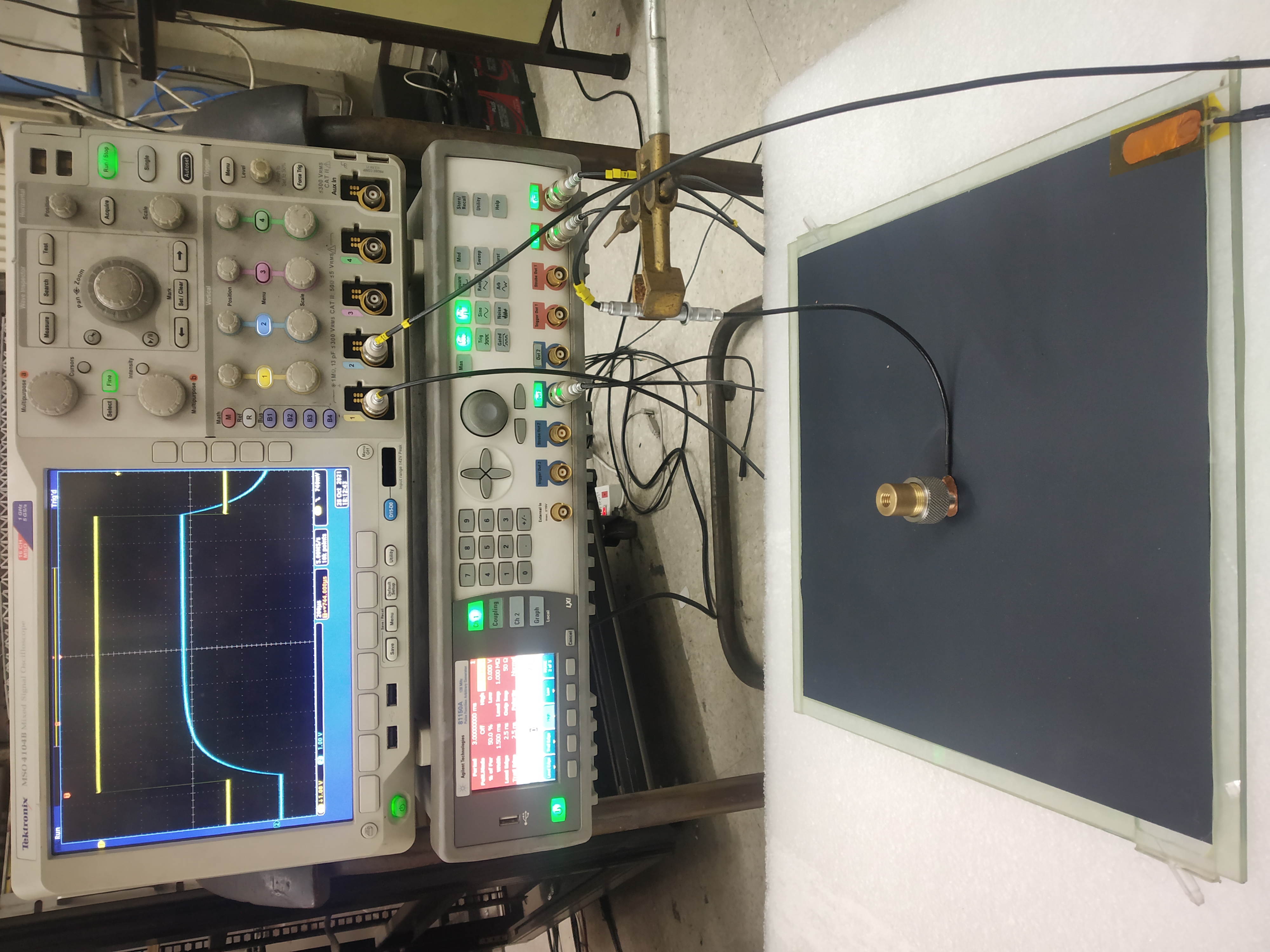}}
		\hspace{10mm}
		\subfloat[Oscilloscope waveforms of potential buildup \label{fig:Meas_timecons_b}]{
			\includegraphics[width=0.4\textwidth]{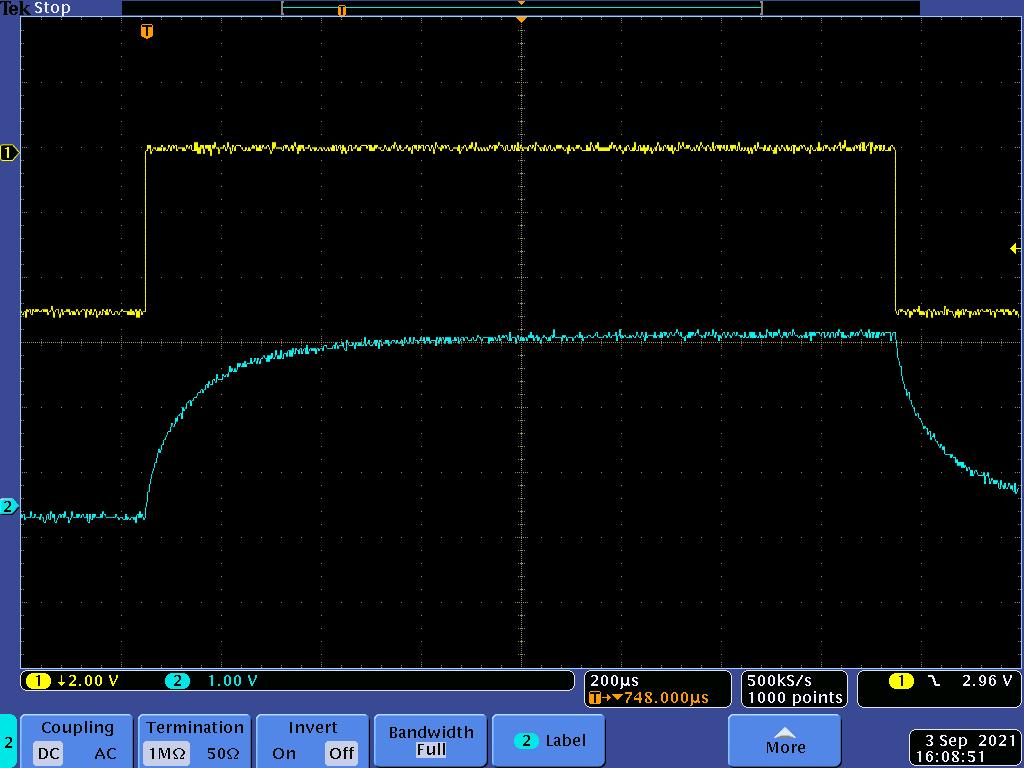}}
	\end{center}
	\caption{\label{fig:Meas_timecons}Measurement of time constant of potential buildup on glass electrode}
\end{figure}

\begin{figure}[h!]
    \centering
    \includegraphics[width=0.8\textwidth]{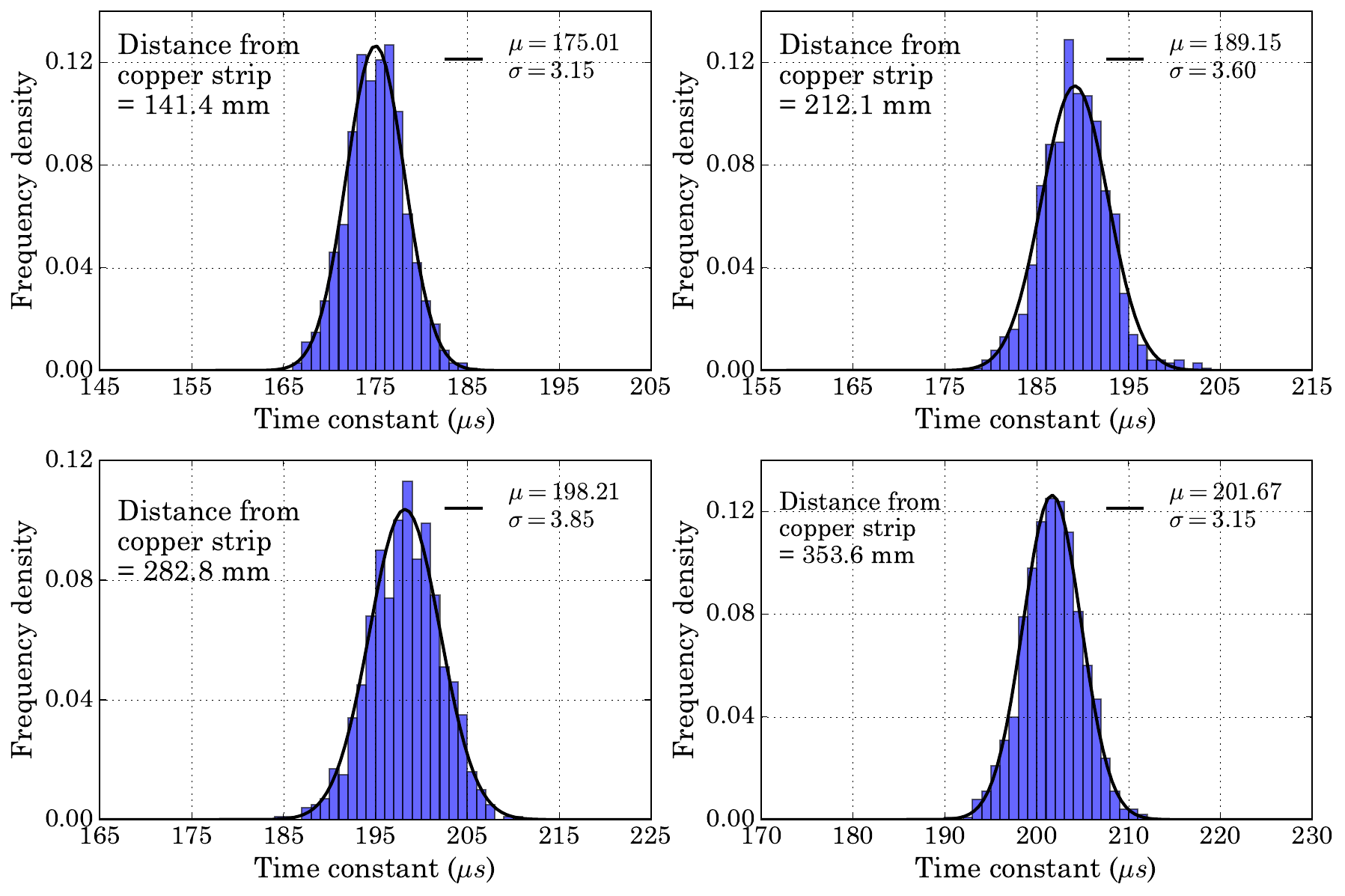}
    \caption{Histogram of time constant measured at four specific locations}
    \label{fig:Hist_timecons}
\end{figure}

It is obvious that for the resistive electrode, there should be a time constant for building up of the potential.
It can be defined as the time elapsed to attain a factor of (1 - 1/$e$) of the applied voltage by the electrode.
The time constant has been evaluated for variation in different electrical and physical parameters of the device components to study their effect on it.
The numerical model has been tested by comparing its results to the measured data obtained for the top electrode of the glass RPC. 
An image of the experimental setup has been displayed in figure \ref{fig:Meas_timecons_a}.
Using a function generator (\textit{model: Agilent 81150A}), a square pulse of amplitude $5~V$ has been supplied to the graphite coat through the copper tape.
The response at the probe has been measured on an oscilloscope (\textit{model: Tektronix MSO4104B}).
The corresponding waveform (in blue) has been depicted in figure \ref{fig:Meas_timecons_b} along with the input square pulse (in yellow).
The measurement has been carried out at some specific locations along the diagonal connecting the point of high voltage supply.
At each location, the time constant of the potential buildup on the glass electrode has been determined from the mean ($\mu$) of the Gaussian fit to the respective histogram of the measured data, while its standard deviation ($\sigma$) has given the fluctuation in the measurement.
A few typical examples of the time constant histograms produced at four specific locations have been presented in figure \ref{fig:Hist_timecons}.

\begin{figure}[h!]
    \centering
    \includegraphics[width=0.45\textwidth]{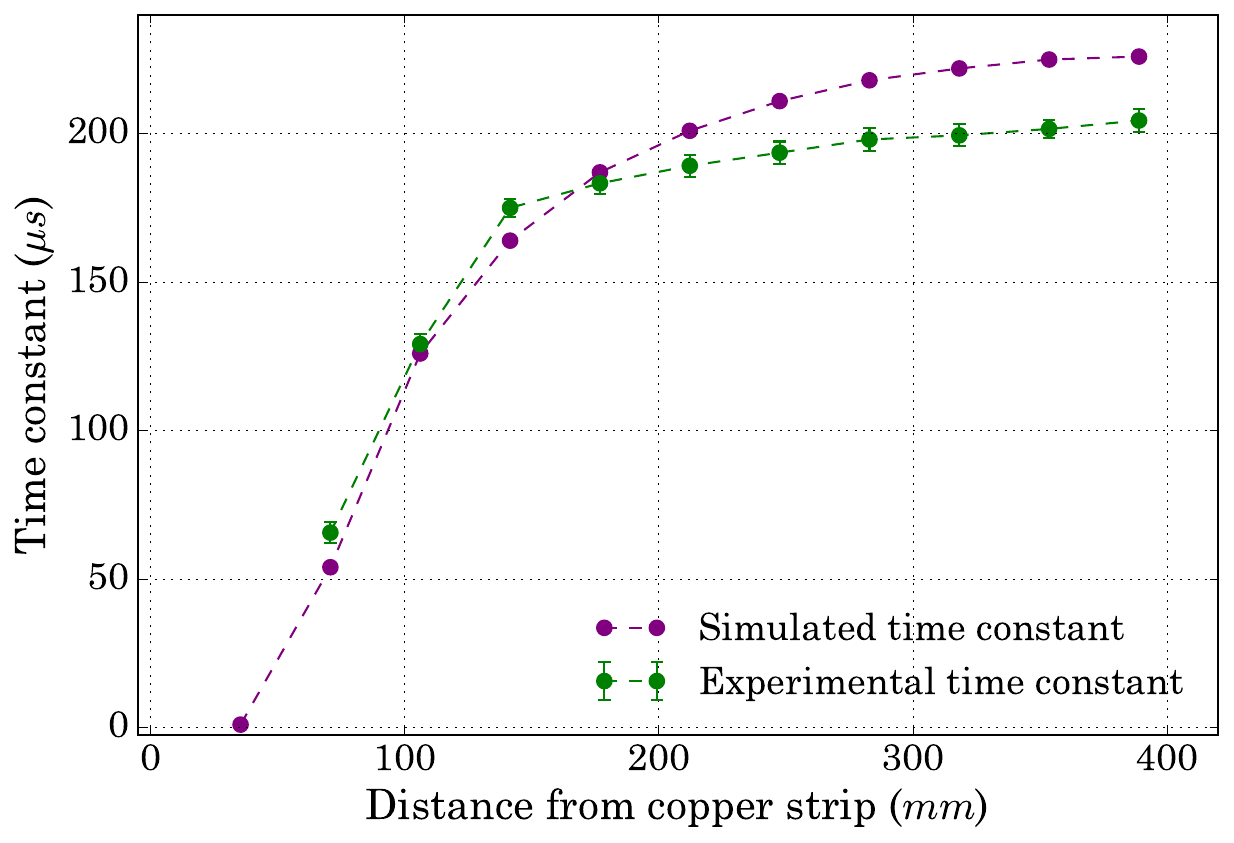}
    \caption{Comparison of measured and simulated time constants of potential buildup on glass RPC}
    \label{fig:Comp_timecons}
\end{figure}

To demonstrate the performance of the numerical model, a comparison of the simulated time constants has been made with the measured data.
It has been displayed in figure \ref{fig:Comp_timecons}.
The numerical model has considered the measured $R_s$-map of the top electrode (as shown in figure \ref{fig:Res_map}) and other physical and electrical parameters as mentioned in table \ref{table3}.
It can be seen from the plot that the time constant for the potential buildup increases with the distance from the point of voltage supply.
There are many parallel resistive and capacitive connections between the contact point and the location of measurement.
The effect of parallel capacitive connection is larger than that of the resistive ones.
This is why time constant increases with the distance as the capacitance grows larger as well. 
The plot exhibits a close agreement between the simulation and experiment.
However, a systematic deviation in the experimental data from the simulated trend has been observed from a distance of $150~mm$ from the voltage contact point.
A possible reason might be a change in the gas gap due to some reason.
An increase in the gas gap should decrease the time constant.
It is worth mentioning here that the uniformity of the gas gap may be investigated by mapping the time constant over the entire electrode.

\begin{figure}[h!]
   \begin{center}
		\subfloat[For different relative permittivity of resistive electrode (glass, Bakelie) \label{fig:Time_elec_a}]{
			\includegraphics[width=0.45\textwidth]{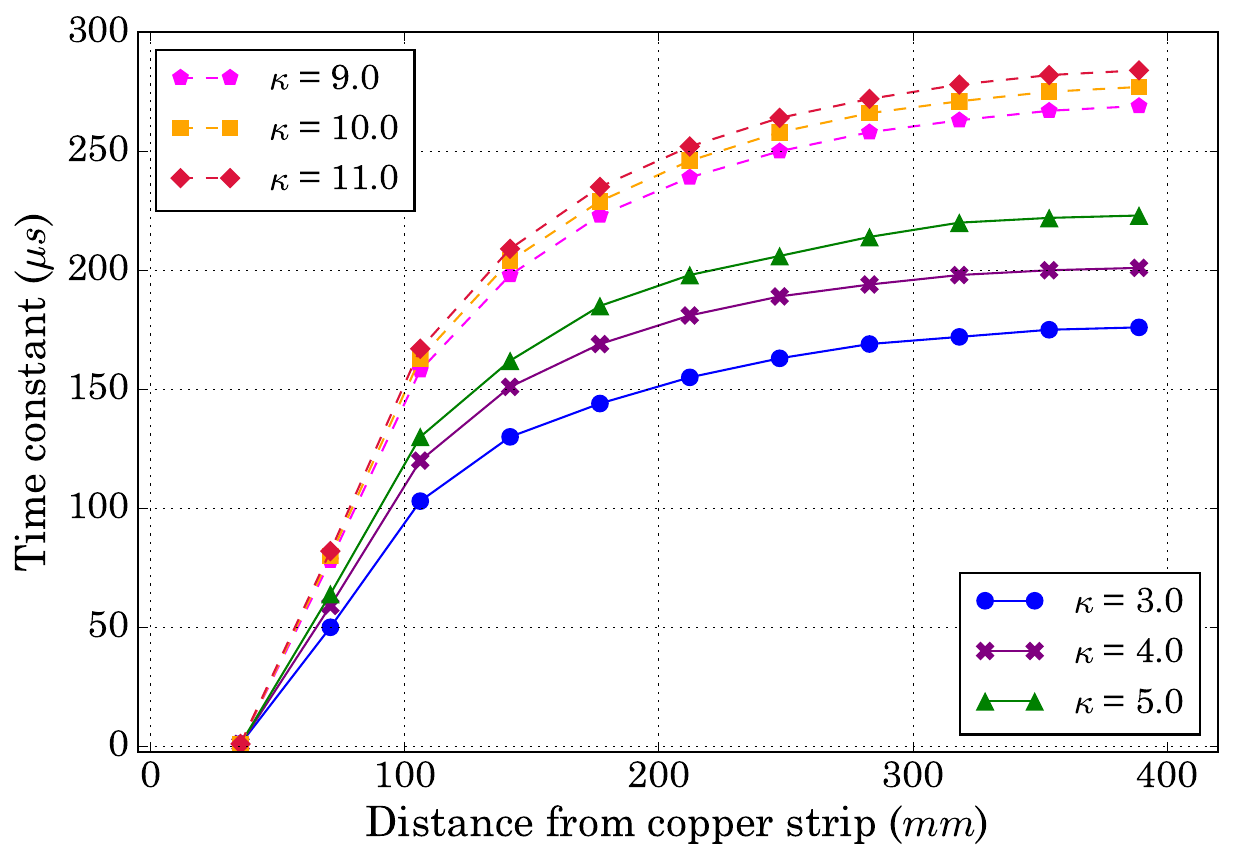}}
		\hspace{2mm}
		\subfloat[For different surface resistivity of coating on resistive electrode (glass, Bakelite) \label{fig:Time_elec_b}]{
			\includegraphics[width=0.45\textwidth]{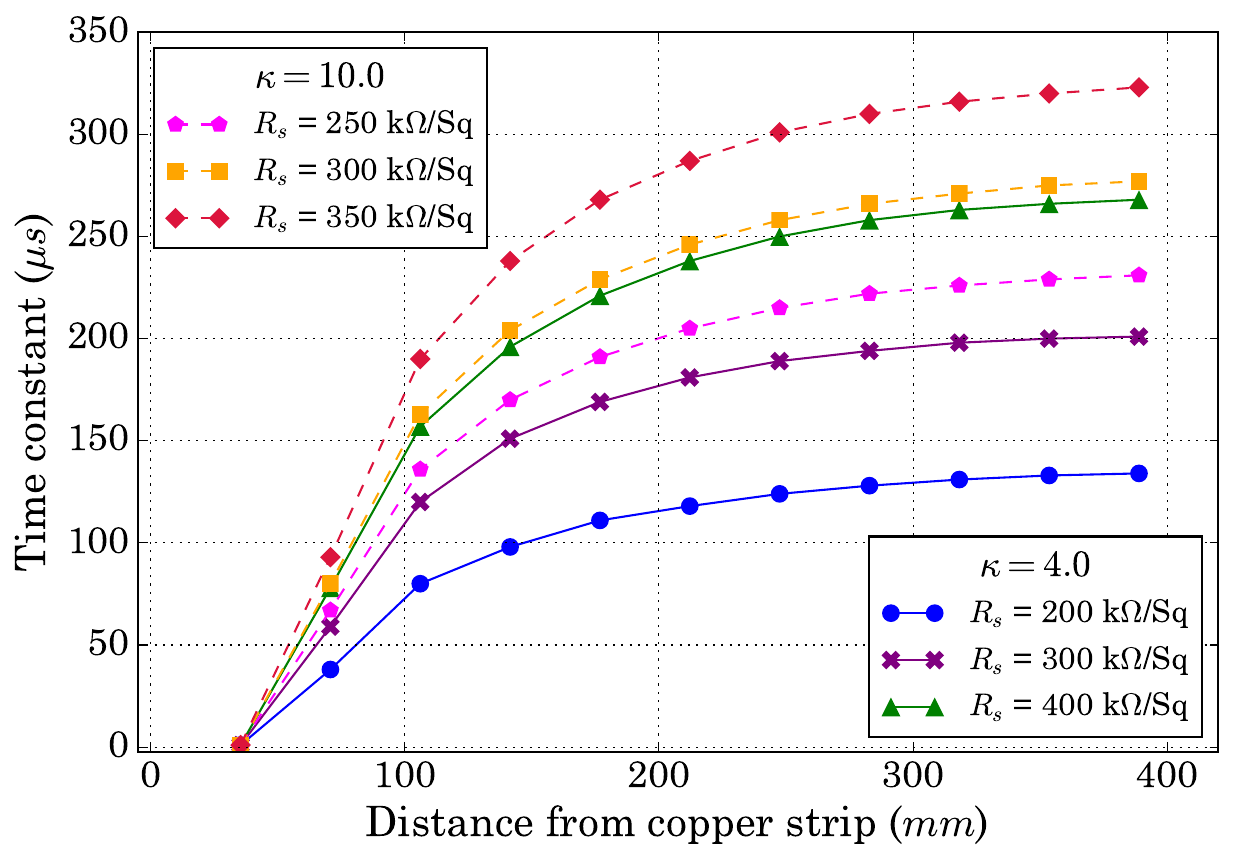}}
	\end{center}
	\caption{\label{fig:Time_elec}Time constant for potential buildup as a function of the distance from the voltage contact point on resistive electrode (glass, Bakelite) for variation in electrical properties of electrode}
\end{figure}

The same effect of increase in the time constant is also expected in case of variation in electrical properties, like relative permittivity and surface resistivity of the electrode, and physical properties, like electrode thickness and gas gap.
In all these case, the capacitive connection overrides the resistive one and hence causes a rise in the time constant with increasing distance. 
The time constant, evaluated as a function of the distance from the voltage contact point for different relative permittivity of the electrode, is shown in figure \ref{fig:Time_elec_a}.
Two common materials, namely, glass ($\kappa = 9, 10, 11$) and Bakelite ($\kappa = 3, 4, 5$), used for building the resistive electrodes, have been considered with a variation in their relative permittivity. 
However, there has been no effect observed in case of variation in the relative permittivity of the spacers.
The reason is that the spacers are too small in the dimension to offer significant capacitance in the device.
The surface resistivity of the resistive electrodes is another electrical property which can govern the time constant.
A calculation of the time constant for different surface resistivity values of two types of electrodes, such as glass ($\kappa = 10.0$ and $R_s = 200, 300, 400~k\SI{}{\ohm\per \sq}$) and Bakelite ($\kappa = 4.0$ and $R_s = 250, 350, 450~k\SI{}{\ohm\per \sq}$), has been performed.
The results have been plotted in  figure \ref{fig:Time_elec_b}.
The plot shows that the time constant is larger for the material with higher relative permittivity for a given surface resistivity.
It rises when the surface resistivity of the electrode increases.

\begin{figure}[h!]
   \begin{center}
		\subfloat[For different electrode thickness\label{fig:Time_phys_a}]{
			\includegraphics[width=0.45\textwidth]{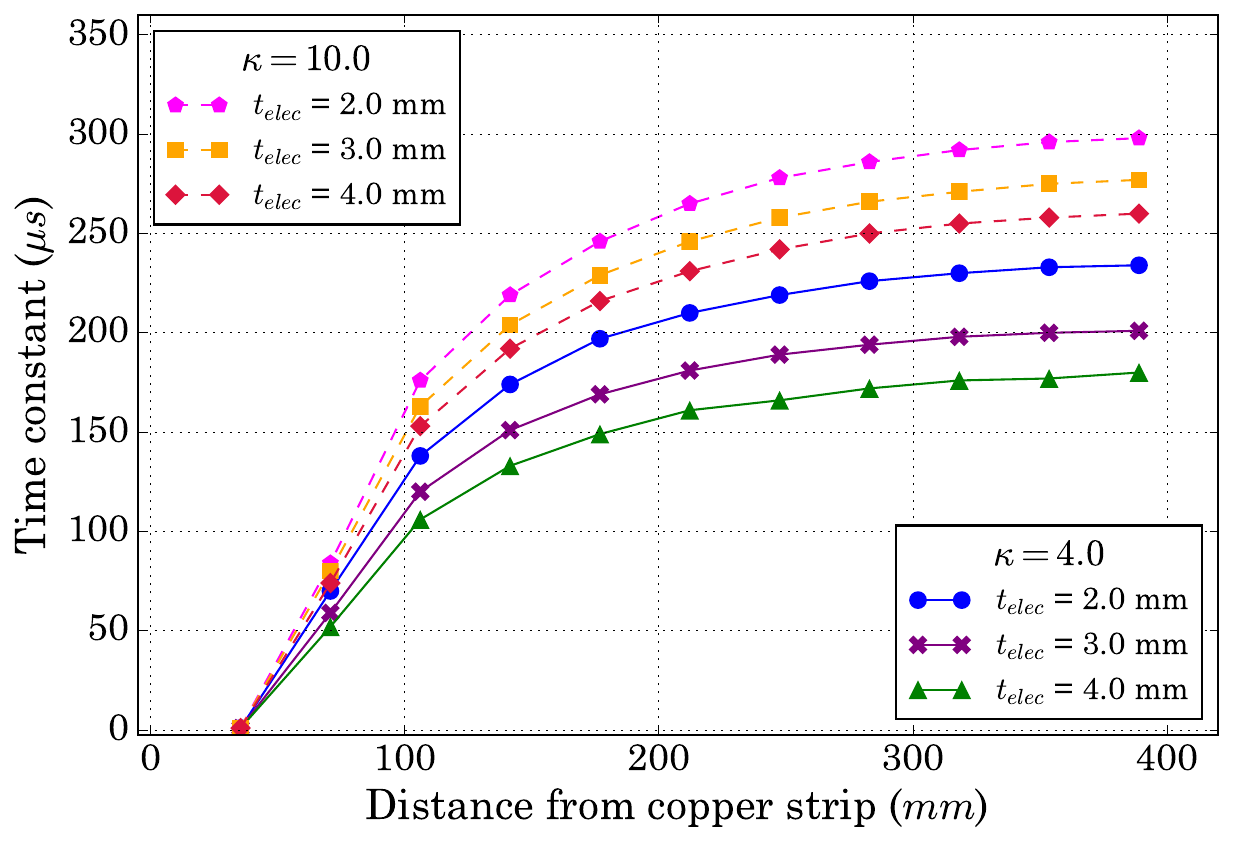}}
		\hspace{2mm}
		\subfloat[For different gas gaps\label{fig:Time_phys_b}]{
			\includegraphics[width=0.45\textwidth]{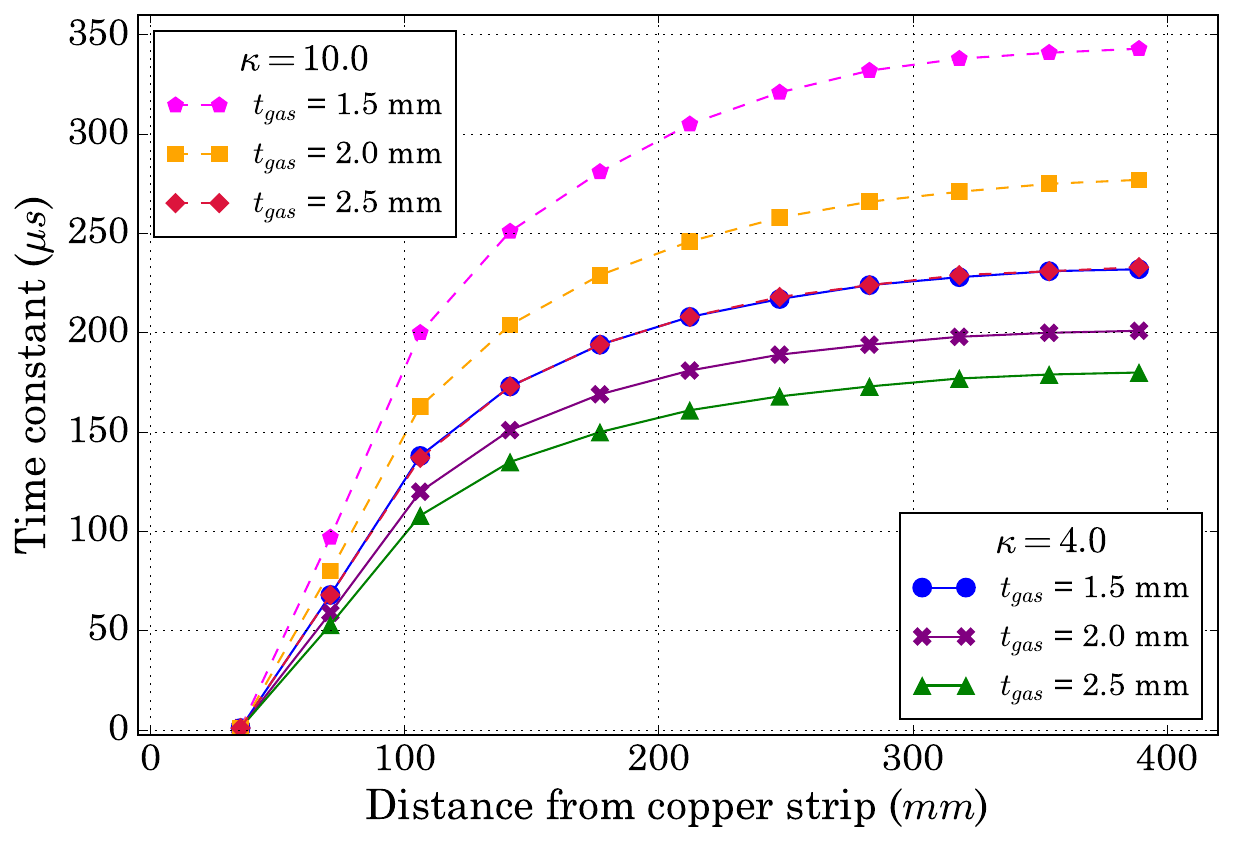}}
	\end{center}
	\caption{\label{fig:Time_phys}Time constant for potential buildup as a function of the distance from the voltage contact point on resistive electrode (glass, Bakelite) for variation in physical properties of the RPC}
\end{figure}

Similarly, the calculated time constants for different values of electrode thickness and gas gap have been shown in figure \ref{fig:Time_phys_a} and \ref{fig:Time_phys_b} for two different electrode materials.
As the change in the thickness of either of the electrode or the gas gap affects the capacitance of the RPC, it changes the time constant in turn.
An increase in the thickness leads to a drop in the capacitance and hence a reduction in the time constant.
As mentioned earlier, this relationship can be utilized for studying the uniformity of electrode thickness or gas gap in the RPC by using a precise experimental setup.

\section{\label{con} Conclusions}
In the present work, a numerical model has been proposed for simulating the properties of the RPC as an electrical system composed of many resistive and capacitive components.
The model has utilized the law of current conservation to evaluate the potential distribution, flow of dark current etc. which are essential to determine the electric field configuration in the RPC.
Being a key factor in envisaging the device performance, the electric field distribution in the RPC requires both qualitative and quantitative understanding, which can be accomplished with the numerical simulation.
The present numerical model has provided us a tool to  study of the electrical properties of the RPC, which should be useful in predicting its performance and analyzing the experimental observations.
It can lead to optimization of design and material for fabricating the RPC components to achieve the desirable performance, if necessary.
Therefore, the work may be found helpful in implementing the RPC in various tracking and triggering experiments.
The results of this model have been compared to another numerical model described in \cite{Ammosov1997} and a close agreement between the models has been observed, which proves the efficacy of the present numerical model in producing reliable field distribution of the RPC with a given design and material configuration.
In our R\&D work about building a muon tomography setup, where large area coverage with uniform response is an essential requirement, the numerical tool has been utilized for studying different design aspects of the RPC and evaluate its performance.
It has been found that a variation of surface resistivity from $100~k\SI{}{\ohm\per \sq} - 100~M\SI{}{\ohm\per \sq}$ can be allowed to achieve the uniform potential distribution, which eventually leads to uniform electric field distribution.
The ratio of the volume resistivity of the spacer to the electrode should be around $10^3$ times or more to sustain the uniform potential over the electrode even at a higher value of its surface resistivity.
This also helps to minimize the distortion in the electric field in the close vicinity of the side and button spacers. 
As the time taken for the potential distribution to get stable is very small with respect to the experimental timescale of muon tomography, the governing factors like thickness of the electrode and gas gap and the relative permittivity can be chosen arbitrarily for this application.
On the other hand, the dependence of this parameter on the thickness of gas gap as well as the electrode gives us a good handle to monitor the uniformity of the gas gap or the electrode thickness.
As increased electrode thickness reduces the dead time, this can be a way to improve the rate capability of the detector.
It can prove to be complementary to the present trend of reducing the thickness of both the gas gap (to reduce the amount of charge) and the electrode (to increase signal induction) to increase the rate capability \cite{Kortner:2021pxe}.
However, further detailed studies are necessary to optimize electrode thickness and gas gap in relation to rate capability.
This information is very important in evaluation of RPC performance as this decides the electric field and in turn the gain of the detector.
	
\acknowledgments
The authors, S. Das and J. Datta are thankful respectively to UGC, Government of India and INO Collaboration for financial support and other helps.
We are grateful to HBNI and SINP for their help and would also like to thank the reviewer for the helpful suggestions.

\bibliography{JHEP}

\end{document}